\documentclass[preprint,11pt,3p]{elsarticle}

\usepackage{amssymb}
\usepackage{amsmath}
\usepackage{graphicx}
\usepackage{hyperref}
\usepackage{mathtools}
\usepackage{mathrsfs}
\usepackage{amsfonts}
\usepackage{bbm}
\usepackage{enumerate}
\usepackage{array}
\usepackage{booktabs}
\usepackage{textcomp}
\usepackage{caption}
\usepackage{cancel}
\usepackage{units}
\usepackage{multirow}
\usepackage{pst-all}
\usepackage{longtable}
\usepackage{relsize}
\usepackage{}[natbib]
\usepackage{amsfonts}
\usepackage[titletoc]{appendix}

\def\eqa{\begin{eqnarray}}
\def\eqae{\end{eqnarray}}
\def\eq{\begin{equation}}
\def\eqe{\end{equation}}
\def\be{\begin{equation}}
\def\ee{\end{equation}}
\def\bea{\begin{eqnarray}}
\def\eea{\end{eqnarray}}
\def\ba{\begin{array}}
\def\ea{\end{array}}
\def\bd{\begin{displaymath}}
\def\ed{\end{displaymath}}

\def\tr{{\rm tr}}
\def\>{\rangle}
\def\<{\langle}

\def\m{\mu}
\def\n{\nu}

\def\q{\theta}

\numberwithin{equation}{section}
\journal{Nuclear Physics B}

\begin{document}

\begin{frontmatter}

\title{N=2 gauge theories on the hemisphere $HS^4$}

\author[label1,label2]{Edi Gava}
\address[label1]{The Abdus Salam International Centre for Theoretical Physics, ICTP\\Strada Costiera 11, 34014 Trieste, Italy}
\address[label2]{INFN, sezione di Trieste, Italy}


\ead{gava@ictp.it}
\author[label1]{Kumar Narain}
\ead{narain@mail.com}
\author[label1]{Nouman Muteeb}
\ead{mmuteeb@ictp.it}
\author[label1]{V.I.Giraldo-Rivera}
\ead{vigirald@gmail.com}

\begin{abstract}
Using localization techniques,  we compute  the path integral  of  $N=2$ SUSY gauge theory coupled to matter  on the hemisphere $HS^4$, with 
either Dirichlet or  Neumann supersymmetric boundary conditions. The resulting quantities are wave-functions of the theory depending on
 the boundary data. The one-loop determinant are computed using $SO(4)$ harmonics basis. We solve kernel and co-kernel equations for the relevant differential operators arising  from gauge and matter localizing actions. 
 The second method utilizes full $SO(5)$ harmonics to reduce the computation to evaluating  $Q_{SUSY}^2$ eigenvalues and its multiplicities.
  In the Dirichlet case,  we show how to glue two wave-functions 
to get back the  partition function of  round $S^4$. We will also  describe how to obtain the same results using $SO(5)$ harmonics basis.
\end{abstract}


\end{frontmatter}
\newpage
\tableofcontents
\newpage
\section{Introduction}
Since the seminal work \cite{Pestun:2007rz} on supersymmetric localization on round $S^4$ there has been an intense activity in the field. This computation was soon generalized to more general curved manifolds in various dimensions \cite{Hama:2012bg,Hosomichi:2012ek,Hama:2011ea} and to various number of supersymmetries  \cite{Nishioka:2014zpa}. Manifolds with boundaries in two and three dimensions were also considered and various quantities computed\cite{Yoshida:2014ssa,Hori:2013ika,Sugishita:2013jca,Honda:2013uca}.\\
For manifolds without boundaries  Atiyah-Singer  Index Theory, together with a clever cohomological reorganization of fields introduced in \cite{Pestun:2007rz},
provides a shortcut to the computation of the one-loop determinants factor and avoids going through the diagonalization of the relevant differential operators
arising from the quadratic part of the localizing action expanded around the saddle points. In particular,
in  \cite{Pestun:2007rz}, it has been shown that the one-loop factor is given by the ratio of determinants of the operator $Q^2$ on the kernel and cokernel of certain differential 
operator determined from the localizing action $QV$, $Q$ being the supercharge entering in the localization procedure and $V$ being some fermionic field. To compute the latter 
determinants one then  uses Atiyah-Bott  fixed point theorem which localizes the computation near the fixed points of $Q^2$ on the base manifold. \\
In this  note  we will  work out the partition function of $N=2$ SUSY  $G$ gauge theory coupled to matter in some $G$-representation $\mathcal R$, on the hemisphere $HS^4$, both for Dirichlet  and Neumann  boundary conditions at the equator of the round $S^4$.
Since applying the above mathematical machinery in the case of manifolds with boundary may be subtle,  we will rather solve the partial differential equations 
which determine the kernel and cokernel of the relevant differential operators , together with the corresponding $Q^2$ quantum numbers and multiplicities.  With this data we write down  the one-loop determinant for both the vector multiplet and  the hypermultiplets, for  the corresponding boundary conditions. This direct approach turns out to be feasible because of 
the symmetry of the problem: we choose the $S^4$ metric with a manifest $SO(4)$ symmetry which is  preserved by the localizing action. This allows to reduce the system of  (coupled) partial differential equations depending on the four coordinates of $S^4$, to a system of ordinary differential equations depending on a radial coordinate $r$. \\
We will then write the expressions for the wave functions on $HS^4$ with Dirichlet and Neumann boundary conditions respectively.  We finally show that the partition function of $N=2$ theory on the round $S^4$ can be obtained by gluing two Dirichlet wave functions corresponding to the two hemispheres $HS^4$s.

Next, for the purpose of completion, we briefly describe how one  can obtain the same results using  spherical harmonics of  the full isometry group  $SO(5)$. The logic is essentially  the same as that for $SO(4)$.

It is interesting to note that the vector multiplet one loop determinant for Dirichlet BCs is not 'half' of that for the full round $S^4$ \cite{Pestun:2007rz}  as one would naively expect. As one solves the zero mode equations for kernel and co-kernel of vector multiplet ,to find the multiplicity of the unpaired bosonic and fermionic modes, it turns out that for the Dirichlet BCs the net multiplicity is one unit less and for the Neumann BCs one unit more that that required for the one loop determinant to be exactly 'half'. This deficiency or access of  one unit of  multiplicity for bulk theory has interesting interpretation in terms of whether one chooses to freeze some degrees of freedom or to add some extra degrees of freedom at the boundary.  \\
The contents in this note are arranged as follows: After a lightening review of $N=2$ SUSY gauge theory coupled to matter in section \ref{susyaction} we move on to compute one-loop determinants of vectormultiplet and hypermultiplet in sections \ref{vectoranalysis} and \ref{matteranalysis}. Everything in the path integral computation is put together in section \ref{wave functions} to get the sought after wave functions of $N=2$ system on hemisphere with Dirichlet and Neumann boundary conditions. Factorization of round sphere $S^4$ $N=2$ partition function is discussed in section \ref{sec:joineds4}. Discussion of $SO(5)$ harmonics and subsequent computation of $Z_{1-loop}^{vec}$ is given in  sections \ref{5harmonics} and \ref{51loop}. A brief set of conclusions are given in section \ref{conclusion}.

\newpage
\section{$N=2$ SUSY Field theory}\label{susyaction}
We will consider $4d$ $N=2$ extended supersymmetric field theories which are defined by eight supercharges in the flat space limit. These supercharges correspond to choosing a pair of chiral, anti-chiral Killing spinors denoted in four component notation by  $\xi$, which satisfies following set of equations.
\bea
D_{\mu}\xi+T_{\nu\rho}\gamma^{\nu\rho}\gamma_\mu\xi=\xi_p,\nonumber\\
\gamma^{\mu}\gamma^{\nu}D_{\mu}D_{\nu}\xi+4D_{\mu}T_{\nu\rho}\gamma^{\nu\rho}\gamma^{\mu}\xi=M \xi
\eea
where $\xi_p\equiv\gamma^{\m}D_{\m}\xi$ and $T_{\mu\nu},M$ are background non-dynamical fields belonging to parent supergravity  theory \cite{Hama:2012bg}. Moreover the covariant derivative $D_{\m}$ also contains a background $SU(2)_R$ R-symmetry gauge field $V^A_{\mu B}$. The values of these auxiliary fields are found, if they exist, for which one is able to solve the Killing spinor equations. This Killing spinor $\xi$ defines the supercharge $Q$ with respect to which we localize the path integral.
\subsection{Physical actions}
The physical actions for vector multiplet  and the matter multiplet  of $N=2$ SUSY  on curved background  are written in \cite{Hama:2012bg}. 
For SYM action we have:
 \begin{align}
 \begin{split}
 L_{YM}&=\text{Tr}\bigg[\frac{1}{2}F^{mn}F_{mn}+16F_{mn}(\bar{\phi}T^{mn}+\phi\bar{T}^{mn})+64\bar{\phi}^2T_{mn}T^{mn}+64\phi^2\bar{T}^{mn}\bar{T}_{mn} \\ & -4D_m\bar{\phi}D^m\phi+2M\phi\bar{\phi}-2i\lambda^A\sigma^mD_m\bar{\lambda}_A-2\lambda^A[\bar{\phi,\lambda_A}]+2\bar{\lambda}^A[\phi,\bar{\lambda}]+4[\phi,\bar{\phi}]^2-\nonumber\\
&\frac{1}{2}D^{AB}D_{AB}\bigg].
 \end{split}
 \end{align}
 When restricted to  the round sphere $S^4$ (appendix [\ref{appendix:harmonics}]), the other supergravity background fields reduce to
 \bea\label{eq:background}
 T_{mn}=0,\quad \bar{T}_{mn}=0,\quad V_{m=0},\quad M=-4.
 \eea
 There are also reality conditions for the fields,  chosen to ensure a well defined path integral. In particular, $\phi=\phi_2+ i \phi_1$ 
 and  $\bar\phi=-\phi_2+ i \phi_1$. The gauginos $\lambda^A$ and $\bar\lambda_A$ 
 of opposite $SO(4)$ chirality carry and $SU(2)_R$ doublet index. $D_{AB}$ is an $SU(2)_R$ triplet of auxiliary fields.
 For non-trivial topological sectors, characterized by instanton number $k$, one has to  add $\theta$-term to the full action:
 \begin{align}
 S_{YM}=\frac{1}{g^2_{YM}}\int d^4x\sqrt{g}L_{YM}+i\frac{\Theta}{8\pi^2}\int \text{Tr}(F\wedge F).
 \end{align}
The action for  for the matter hypermultiplets is:
 \begin{align}
 \begin{split}
  L_{mat}=\frac{1}{2}D_mq^AD^mq_A-q^A\{\phi,\bar{\phi}\}q_A+\frac{i}{2}q^AD_{AB}q^B+\frac{1} {8}(R+M)q^Aq_A-\frac{i}{2}\bar{\psi}\bar{\sigma}^mD_m\psi- \\ \frac{1}{2}\psi\phi\psi+  \frac{1}{2}\bar{\psi}\bar{\phi}\bar{\psi}+\frac{i}{2}\psi\sigma^{kl}T_{kl}\psi-\frac{i}{2}\bar{\psi}\bar{\sigma}^{kl}\bar{T}_{kl}\bar{\psi}-q^A\lambda_A\psi+\bar{\psi}\bar{\lambda}q^A-\frac{1}{2}F^AF_A.
 \end{split}
 \end{align}
where again the supergravity  background satisfies eq.( \ref{eq:background}) and a with proper reality conditions for the fields is  understood. The scalars $q^A$
carry an  $SU(2)_R$ index $A$ and in addition an index $I=1,...,2 q$ of  a symplectic representation of the gauge symmetry group  $G\subset Sp(q)$,  therefore
it is possible to impose a reality condition on them in the usual way.  The spinors $\psi$ carry index $I$  and  $F^A$ are auxiliary fields.
\subsection{Localizing actions}

The localization technique proceeds first by identifying a supercharge $\hat{\textbf{Q}}$  and then a $\hat{\textbf{Q}}$-exact localizing action,
with positive definite bosonic part, by which one  perturbs the physical action in such a way that the path integral is independent
of the perturbation. One then shows that the path integral localizes at the supersymmetric saddle
points of the localizing action, in the sense that the one-loop approximation around them is exact.

The localizing supercharge $\hat{\textbf{Q}}$   depends on a choice of Killing spinors, $ \xi^{A\alpha}$ and  $\bar\xi^{A\dot{\alpha}}$, which we arrange in a four-by-two matrix
using four component $SO(4)$ spinors which are also $SU(2)_R$ doublets. Killing spinor $\xi$ is taken as Grassmann-even and $\hat{\textbf{Q}}$  is Grassmann-odd.  With the background metric:
\bea
ds^2=g_{\mu\nu}dx^{\mu}dx^{\nu}=dr^2+\frac{\sin(r)^2}{4}\big(d\theta^2+\sin\theta^2d\phi^2+(d\psi+\cos\theta d\phi)^2\big).
\eea
which can be written in terms of $SU(2)$ left-invariant one-forms, a solution of the $N=2$  Killing spinor equations is given by:

\bea
\xi=\left(
\begin{array}{cc}
 \frac{\cos \left(\frac{r}{2}\right)}{\sqrt{2}} & 0 \\
 0 & \frac{\cos \left(\frac{r}{2}\right)}{\sqrt{2}} \\
 \frac{i \sin \left(\frac{r}{2}\right)}{\sqrt{2}} & 0 \\
 0 & -\frac{i \sin \left(\frac{r}{2}\right)}{\sqrt{2}} \\
\end{array}
\right).
\eea

with the index structure $\xi\equiv(\xi_{\alpha A},\bar{\xi}_{\dot{\alpha}A})$ where $\alpha,\dot{\alpha}$ are Lorentz indices and  $A$ is R symmetry index\footnote{In four component notation we use the matrix $\left(
\begin{array}{cccc}
 0 & 1 & 0 & 0 \\
 -1 & 0 & 0 & 0 \\
 0 & 0 & 0 & 1 \\
 0 & 0 & -1 & 0 \\
\end{array}
\right)$ to raise and  and lower the $\alpha,\dot{\alpha}$ indices}.
And it is normalized to  $\xi^A\xi_A+\bar{\xi}_A\bar{\xi}^A=1$. These Killing spinors give rise to a Killing
vector $v=2\frac{\partial}{\partial \psi}$. The corresponding supercharge $ \hat{\textbf{Q}}$ squares on all
fields to bosonic symmetry generators:
\bea
\hat{\textbf{Q}}^2=\mathcal{L}_v + \rm{R} + \rm{Gauge}_\Lambda.
\eea
where $\mathcal{L}_v$ is the Lie derivative along $v$,  R  and $\rm{Gauge}_\Lambda$ are, respectively, R-symmetry and field dependent 
gauge transformation parameter $\Lambda$:
\bea\label{eq:gauge}
\Lambda=-v^\mu A_\mu-2 i\phi_1-2 \cos(r) \phi_2.
\eea
The localizing action, $S_{loc}= \hat{\textbf{Q}} V$\footnote{ $\hat{\textbf{Q}} $ should be defined by including the BRST component
and correspondingly  $V$ should include the ghost part, as it will be shown in the following.},  is determined by the fermionic field $V$, for which,
a convenient expression in terms of  original fields is:
\begin{align}
\label{eq:QV1}
V=\text{Tr}[(\hat{\textbf{Q}}\lambda_{\alpha A})^{\dagger}\lambda_{\alpha A}+(\hat{\textbf{Q}}\bar{\lambda}^{\dot{\alpha}}_A)^{\dagger}\bar{\lambda}^{\dot{\alpha}}_A],
\end{align}
where $\dagger$ means complex conjugation. One can show that the localization locus is given by
\begin{align}\label{eq:locusvector}
 F_{\mu\nu}=0, \qquad \phi=\bar{\phi}= a_0,\qquad D_{AB}=-i a_0 \omega_{AB},
\end{align}
where $\omega_{AB}=-8( \xi_A\xi_{pB}+\bar{\xi}_B\bar{\xi}_{pA})$
and $\xi_p=\gamma^{\m}D_{\m}\xi$. Therefore $\phi_2=0$ and $\phi_1=a_0$,
$a_0$  being a  constant  element of the Lie algebra of the gauge group $G$. 
Note that at the saddle point, choosing $A_\mu=0$, $\Lambda$ reduces to a constant gauge parameter,
 $\Lambda=-2i a_0$, given by the v.e.v. of the scalar  $\phi_1$.\\
Here we stress that although $A_\mu$ is a pure gauge, due to non-trivial $\pi_3(SU(2))\simeq\pi_3(S^3)\simeq \mathbb{Z}$ of the boundary there are equivalence classes of gauge connections characterized by integer $k$. And to compute perturbative part the quadratic fluctuations will be  in general around vacuum labelled by $k$. \\
As mentioned in the introduction, it is convenient to introduce new fermionic fields which are linear combinations of the gauginos and carry integer spins, which are part of the cohomological   fields,
 \bea\label{eq:cohomv}
\psi=\hat{\textbf{Q}}\phi_2,\quad \psi_\mu=\hat{\textbf{Q}}A_\mu,\quad \vec{\chi}=\xi^{A}\lambda_{B}\vec{\sigma}_A^B.
 \eea
The main point about introducing these fields is that, after gauge fixing and introducing the ghost-antighost system $c,\tilde{c}$,  and extending 
$\hat{\textbf{Q}}$ to include the BRST generator, the bosonic fields $X_0=(A_\mu, \phi_2)$
and the fermionic ones, $X_1=(\vec{\chi},c,\bar{c})$, are lowest components of $\hat{\textbf{Q}}$ multiplets \footnote{Here we neglect the ghost-for-ghost system,
which takes care of the $c$,$\tilde c$ zero modes}. 
One can rewrite \ref{eq:QV1}  in terms of cohomological  variables \cite{Hama:2012bg}, however after some linear algebra, one can show that the
relevant super determinant corresponding to the quadratic part of $\hat{\textbf{Q}}V$ can be expressed in terms of the super determinant of the $\hat{\textbf{Q}}^2$
operator on the spaces $X_0$ and $X_1$. Furthermore, since these spaces are related by a differential operator $D_{10}$ which commutes with  $\hat{\textbf{Q}}V$ 
and can be read off from $V$, at the end the super determinant, upto an overall  sign ambiguity, reduces to
\bea\label{eq:z1loop}
Z_{1-loop}=(\frac{\textbf{det}_{CokerD_{10}}\hat{\textbf{Q}}^2}{\textbf{det}_{KerD_{10}}\hat{\textbf{Q}}^2})^{\frac{1}{2}}.
\eea
that is, it is enough to compute the spectrum of $\hat{\textbf{Q}}^2$ on the kernel and cokernel of $D_{10}$ , where the latter equals the kernel  $D_{10}^{\dagger}$.
The differential  operator $D_{10}$ is identified 
from the terms in $V$ which are  bilinear in $X_0$ and $X_1$, after expanding around the saddle point field configuration.
The terms relevant to our analysis are the following
\bea
V_{vec}=e\text{Tr}\bigg(\chi_a ({\hat{\textbf{Q}}\chi}_a)^\dagger+\frac{1}{e}\text{c} D_{\nu}(\text{e}{\hat{\textbf{Q}}\psi_{\nu}})^\dagger+\frac{1}{e} \bar{c} D_{\nu}(\text{e}A^{\nu} )\bigg).
\eea
Here $e=\sqrt{g}$.

Similarly, for the matter part the localizing action is obtained from:
\begin{align}
 V_{mat}=e \text{Tr}[(\hat{\textbf{Q}}\psi_{\alpha I})^{\dagger}\psi_{\alpha I}+(\hat{\textbf{Q}}\bar{\psi}^{\dot{\alpha}}_I)^{\dagger}\bar{\psi}^{\dot{\alpha}}_I].
\end{align}
and a trivial localization locus
\begin{align}\label{eq:locusmatter}
 q_{IA}=0,\qquad F_{IA}=0.
 \end{align}
As for the vector case, it is convenient to change fermionic variables from $\psi_{\alpha}$ to $\Sigma_A=\hat{\xi}_A\psi$, where  $\hat{\xi}$ 
is a spinor orthogonal to $\xi$. The cohomological fields are $X_0=q_A$ and $X_1=\Sigma_A$ and the same linear algebra argument as
before shows that the matter one-loop contribution is given by the superdeterminant of $ \hat{Q}^2$ on the kernel and cokernel of the  operator
$D_{10}^m$ mapping $X_0$ to $X_1$, which can be read off from  $V_{mat}$ by keeping the terms bilinear in $q$ and $\Sigma$.
\subsection{Main content of computation}
The main content of this work is the computation of the partition function (actually,  wave-function ) of $N=2$ supersymmetric gauge theory on a hemisphere $HS^4$ ,  with supersymmetric boundary conditions. The relevant one-loop determinants for gauge and matter multiplets are computed by
 direct analysis of partial differential equations defining the kernel and cokernel of $D_{10}$  differential operators. In other words , we take the differential operator $D_{10}$  
and its adjoint counterpart $(D_{10})^{\dagger}$  from the fermionic functional $V$ and then solve the zero mode partial differential equations for them. We explicitly find the solutions and their multiplicities by diagonalizing the differential equations by expanding fields in  $SO(4)\sim SU(2)_L\times SU(2)_R$ harmonics. The kernel equations
for $X_0$ fields are obtained by varying $V_{loc}$ with respect to  $X_1$ fields and vice versa of the cokernel equations. We write them in the appendix. The differential equations can
be expressed in terms  of $SU(2)_L$ generators and $r$-derivatives, so that $SU(2)_R$ is spectator and provides  the multiplicities, once we solve the ordinary  differential equations
in the $r$ coordinate.\\
We work with the round $S^4$  and then we adapt it  to the case of  hemisphere $HS^4$,  where we impose appropriate boundary
conditions  to the zero modes of $D_{10}$ and $(D_{10})^{\dagger}$ at the equator of $S^4$, taken to be at $r=\pi/2$.  
As it turns out the solution set for the zero modes of  $D_{10}$ is empty and those of $(D_{10})^{\dagger}$ is non-empty for the vector multiplet, whereas the converse is true for the matter multiplet.
The  $\hat{\textbf{Q}}^2$ eigenvalues and their multiplicities will  then give us, by  using eq.(\ref{eq:z1loop}),  the expression for the determinants.\\
The analysis for kernel and cokernel equations respectively in sections  \ref{vectoranalysis} and \ref{matteranalysis} is done for $U(1)$ gauge group. The generalization for a  non-abelian gauge group $G$  is straightforward. For vector multiplet we  have to multiply the index by the character $\sum_{\alpha \in Roots}e^{i \alpha.a}$ of adjoint representation of G and for matter multiplet by the character $\sum_{\rho \in R}e^{i \rho.a}$ in the representation $R$ of G.\\
Some comments on BRST analysis are in order. The standard covariant  way to fix the gauge redundancy of the action is  BRST formalism. In an abelian gauge theory like $U(1)$ the BRST charge $Q_B$ which parametrized the gauge freedom is nilpotent. To generalize it to non-abelian gauge group the corresponding BRST charge $Q_B$ squares to a constant gauge transformation $a$ that is ultimately identified with the zero mode of the scalar field as a solution to localization equations. This constant gauge parameter $a$ is integrated over to get the partition function on $S^4$.

\section{Vector-multiplet contribution}\label{vectoranalysis}
Perturbative part of  partition function corresponds to computing one loop determinant of  the quadratic fluctuations around classical field configuration given by the saddle point solutions (\ref{eq:locusvector})(\ref{eq:locusmatter}). The differential operator $D_{10}$ from which we get kernel and cokernel equations, contains only the fluctuating part of the quantum fields. In the vector multiplet the fluctuating part of only $\phi_2\in X_0$ is relevant, whereas $\phi_1$ and $D_{AB}$ contributes only classically. Hence in our discussion of kernel equations we will set $\phi_1=0,D_{AB}=0$. As detailed in the appendix, it is convenient to work with tangent space basis for the gauge fields $A_\mu$ :
 for the $S^3$ directions  $\mu=1,2,3$, the flat basis is $A_a=l^{a\mu}A_{\mu}$ where $ a=1,2,3$ or $a=+,-,3$ where $A_+\equiv A_1+i A_2, A_-\equiv A_1-i A_2$ 
 Similarly for the fermionic fields in $X_1$ we choose a complex basis  $\chi_+\equiv\chi_1+i \chi_2, \chi_-\equiv\chi_1-i \chi_2$.
 The first thing to observe is that, since $D_{10}$ commutes with $\hat{\textbf{Q}}^2$, it will close in $X_0$ on fields of the same  $\hat{\textbf{Q}}^2$ eigenvalue and similarly
 for   $D_{10}^\dagger $ on $X_1$. As we will see in the following sections that $D_{10}$ and $D_{10}^{\dagger}$  will only involve $SU(2)_L$ operators, so all the  fields in $X_0$ will have the same $SU(2)_R$ weight denoted here by $q_R$ and all the fields in $X_1$ will carry weight $-q_R$. The first data  to compute are the  $\hat{\textbf{Q}}^2$ eigenvalues on the cohomological fields. We start with the gauge fields in $X_0$:
 \bea\label{eq:q2v}
\hat{\textbf{Q}}^2A_3&=&2\partial_{\psi}A_3,\qquad\quad \hat{\textbf{Q}}^2A_4=2\partial_{\psi}A_4,\nonumber\\
\hat{\textbf{Q}}^2A_+&=&2( \partial_{\psi}-i)A_+,\quad   \hat{\textbf{Q}}^2A_-=2( \partial_{\psi}+i)A_-.
\eea
and $ \hat{\textbf{Q}}^2\phi_2=2\partial_{\psi}\phi_2$. Notice the shifts in $A_\pm$.
As for the fields in $X_1$ we have:
\bea\label{eq:q2m}
\hat{\textbf{Q}}^2\chi_3&=&2\partial_{\psi}\chi_3,\quad \hat{\textbf{Q}}^2c=-i\partial_{\psi}c,\quad \hat{\textbf{Q}}^2\bar{c}=2\partial_{\psi}\bar{c},\nonumber\\
\hat{\textbf{Q}}^2 \chi_+&=&2( \partial_{\psi}-i)\chi_+,\quad   \hat{\textbf{Q}}^2\chi_-=2( \partial_{\psi}+i)\chi_-.
\eea
Again notice here the shifts in $\chi_\pm$.\\

It is then easy to see that, Fourier transforming every field $\Phi$  in $V_{loc}$ as  $e^{i (q_L\psi+q_R\phi)}\Phi(\theta,r)$ , the pairing of $X_0$ and $X_1$ 
involves only terms with net $q_R=0$ and net $\hat{\textbf{Q}}^2=0$ and the $\phi$, $\psi$ dependence drops out. In this way  partial differential equations in four variables are converted into
partial differential equations  in terms of functions of two variables $(\theta,r)$. These functions are the Fourier coefficients depending on $SU(2)_L\times SU(2)_R$  charges 
$(q_L,q_R)$. Moreover in tangent space basis kernel and cokernel equations can be written in terms of  $SU(2)_L$ generators with an inert $SU(2)_R$ charge $q_R$.

\subsection{Kernel equations}
The fact that   kernel as well as the cokernel differential equations can be written in terms of $SU(2)_L$ generators $J_L$ only, 
implies in  particular that  $SU(2)_R$ commutes with the equations,  and therefore the solutions will organize in 
$SU(2)_R$ multiplets of
dimensions $2 j_R+1$, where the possible values of $j_R$ are $j_L\pm1, j_L$ , depending on the spherical harmonics involved,  as it is detailed in the appendix \ref{appendix:harmonics}. 
In more detail, if one introduces the expressions:
\bea
J^{-}A_+&=&e^{-i q_L\psi}e^{-i q_R\phi}l^{-\mu}\partial_{\mu}(e^{i (q_L+1)\psi}e^{i (q_R)\phi}A_+(\theta,r)),\nonumber\\
J^{+}A_-&=&e^{-i q_L\psi}e^{-i q_R\phi}l^{+\mu}\partial_{\mu}(e^{i (q_L-1)\psi}e^{i (q_R)\phi}A_-(\theta,r)),
\nonumber\\
J^{-}A_3&=&e^{-i (q_L-1)\psi}e^{-i q_R\phi}l^{-\mu}\partial_{\mu}(e^{i q_L\psi}e^{i (q_R)\phi}A_3(\theta,r)),\nonumber\\
J^{+}A_3&=&e^{-i (q_L+1)\psi}e^{-i q_R\phi}l^{+\mu}\partial_{\mu}(e^{i q_L\psi}e^{i (q_R)\phi}A_3(\theta,r)),
\nonumber\\
J^{-}A_4&=&e^{-i (q_L-1)\psi}e^{-i q_R\phi}l^{-\mu}\partial_{\mu}(e^{i q_L\psi}e^{i (q_R)\phi}A_4(\theta,r)),\nonumber\\
J^{+}A_4&=&e^{-i (q_L+1)\psi}e^{-i q_R\phi}l^{+\mu}\partial_{\mu}(e^{i q_L\psi}e^{i (q_R)\phi}A_4(\theta,r)),
\nonumber\\
J^{+}\Lambda&=&e^{-i (q_L+1)\psi}e^{-i q_R\phi}l^{+\mu}\partial_{\mu}(e^{i q_L\psi}e^{i (q_R)\phi}\Lambda(\theta,r)).
\eea
one can show that the kernel equations can be written in the following form:
\bea\label{eq:kernelequations}
\mathcal{E}_1&=&\frac{i}{4}\tan(r)\sin(\theta)\bigg((J^{-}A_3+\frac{1}{4}\sin(2 r)J^{-}A_4)+
\frac{1}{2}\cos(r)(2 q_LA_-(\theta,r)\cos(r),\nonumber\\
&-&\sin(r)\partial_rA_-(\theta,r) \bigg)\nonumber\\
\mathcal{E}_2&=&\frac{i}{4}\tan(r)\sin(\theta)\bigg((-J^{+}A_3+\frac{1}{4}\sin(2 r)J^{+}A_4)-
\frac{1}{2}\cos(r)(2 q_LA_+(\theta,r)\cos(r)+\nonumber\\
&x&\sin(r)\partial_rA_+(\theta,r) \bigg),\nonumber\\
\mathcal{E}_3&=&\frac{i}{8}\sec(r)\sin(\theta)\bigg( \sin(2 r) (-J^-A_++J^+A_-)-(2 q_L A_4(\theta,r)\cos(r)^2\sin(r)^2\nonumber\\
&+&\tan(r)(A_3(\theta,r)(3+\cos(2 r))+\sin(2 r)\partial_r A_3(\theta,r)))
\bigg),\nonumber\\
\mathcal{E}_4&=&\frac{1}{8}\sin(r)\sin(\theta)\bigg( 2(J^-A_++J^+A_-)-(4 q_L A_3(\theta,r)+\sin(r)(3 A_4(\theta,r)\cos(r)
\nonumber\\
&+&\sin(r)\partial_rA_4(\theta,r)))
\bigg),\nonumber\\
\mathcal{E}_5&=&\frac{1}{2} \sin (\theta ) \sin (r) \left(-4 \partial_{\theta}^2\Lambda(\theta ,r)-\sin ^2(r) \partial_r^2\Lambda(\theta ,r)-4 \cot (\theta )
   \partial_{\theta}\Lambda(\theta ,r)\right)\nonumber\\
   &-&\left(3 \sin (r) \cos (r) \partial_r\Lambda(\theta ,r)+4 \csc ^2(\theta ) \Lambda(\theta ,r) \left(\text{q}_L^2-2
   \text{q}_L \text{q}_R \cos (\theta )+\text{q}_R^2\right)\right).
\eea
Where, in the last equation we have traded $\phi_2$ with $\Lambda$ , for fluctuating part of  $\phi_1=0$, in order to simplify the equation.
Note that $\mathcal{E}_5=0$ can be written as Laplacian acting of $\Lambda$
\bea
\nabla^2\Lambda=0.
\eea
This equation has no smooth solution on  $S^4$ apart from the constant and since we will be considering here in $j_L>0$, we set it to 0.
Whereas on hemisphere $HS^4$ with boundary there is constant function as solution for the special value $j_L=0$.  But if we impose supersymmetric boundary conditions this constant solution must be set to zero. This is discussed briefly in appendix. Therefore $\mathcal{E}_5$ drops from our further discussions. \\
The coefficient $\tan(r)$ or $\sec(r)$ in kernel equations (\ref{eq:kernelequations}) may appear problematic because it blows up at $r=\frac{\pi}{2}$. However  it is  only an artifact of the redefinition  of $\phi_1(\theta,r)$ in favor of $\Lambda(\theta,r)$
to simplify the form of  differential equations. On the other hand if redefine $A_3(\theta,r)$ instead of $\phi_1(\theta,r)$
\bea
\text{A}_3(\theta ,r)=-\frac{1}{2}  (2 i\phi_1(\theta ,r)+2 \cos (r) \phi_2(\theta ,r)+\Lambda(\theta ,r))
\eea
we will get kernel equations which will be well defined at $r=\frac{\pi}{2}$.
The only change in the kernel equations will be that $A_3(\theta,r)$ is replaced by $\phi_1(\theta,r)$ but the rest of the analysis will remain the same resulting in the same eigenvalue spectrum and degeneracies.
A similar argument holds for the cokernel  equations.\\
The following modes carry the same value of $\hat{\textbf{Q}}^2$, taking into account the shifts in (\ref{eq:q2v}):
\bea
A_+(\theta,r)&=&Y^{(j_L,q_L+1,q_R)}(\theta)A_+^{(j_L,q_L+1,q_R)}(r),\qquad
A_-(\theta,r)=Y^{(j_L,q_L-1,q_R)}(\theta)A_-^{(j_L,q_L-1,q_R)}(r),\nonumber\\
A_3(\theta,r)&=&Y^{(j_L,q_L,q_R)}(\theta)A_3^{(j_L,q_L,q_R)}(r),\qquad\qquad
A_4(\theta,r)=Y^{(j_L,q_L,q_R)}(\theta)A_4^{(j_L,q_L,q_R)}(r).\nonumber\\
\eea
where the $Y$ functions are the $\theta$ dependent part of the scalar spherical harmonics with the indicated quantum numbers. 
Now we analyze the kernel equations separately for all the possible values of  $q_L$ for which we may get a non-trivial solution.\\
i)$q_L=j_L+1,-j_L-1$\\
The kernel equations evaluate to
\bea
\mathcal{E}_1&=&0,\qquad \mathcal{E}_2=0,\qquad \mathcal{E}_3=0,\nonumber\\
\mathcal{E}_4&=&-\frac{i}{8}\sin(r)\sin(\theta)Y^{(j_L,-l_L,-q_R)}(\theta)(2(1+j_L)\cos(r)A_+^{(j_L,-l_L,-q_R)}(r)\nonumber\\
&+&\sin(r)\partial_rA_+^{(j_L,-l_L,-q_R)}(r)).
\eea
Solving the differential equation $\mathcal{E}_4=0$ we get the solution
\bea
A_+^{(j_L,-l_L,-q_R)}(r)=A^0_+\sin(r)^{-2(1+j_L)}.
\eea
This is clearly a singular solution at the two poles of round $S^4$. So we have to set $A^0_+=0$. If there is a physical boundary at $r=\frac{\pi}{2}$ the result does not change.\\
For $ q_L=-j_L-1$:
\bea
\mathcal{E}_1&=&0,\qquad \mathcal{E}_2=0,\qquad \mathcal{E}_4=0,\nonumber\\
\mathcal{E}_3&=&-\frac{i}{8}\sin(r)\sin(\theta)Y^{(j_L,j_L,-q_R)}(\theta)(2(1+j_L)\cos(r)A_+^{(j_L,j_L,-q_R)}(r)\nonumber\\
&+&\sin(r)\partial_rA_+^{(j_L,j_L,-q_R)}(r)).
\eea
Here $\mathcal{E}_3$ can be solved to give
\bea
A_+^{(j_L,j_L,-q_R)}(r)=A^0_+\sin(r)^{-2(1+j_L)}.
\eea
which is again a singular solution and $A^0_+=0$.\\
ii)$q_L=+j_L,-j_L$\\
First for $q_L=j_L$ solving $\mathcal{E}_1=0$ gives the solution
\bea
A_+^{(j_L,1-j_L,-q_R)}(r)&=&-\frac{1}{2}\cos(r)((3+\cos(2 r))\sec(r)^3A_3^{(j_L,-j_L,-q_R)}(r)+
+2(j_L\sin(r)A_4^{(j_L,-j_L,-q_R)}(r)\nonumber\\
&+&\sec(r)\tan(r)\partial_rA_3^{(j_L,-j_L,-q_R)}))).
\eea
for $\mathcal{E}_2=0$ 
\bea
A_3^{(j_L,-j_L,-q_R)}(r)=A_3^{0}\cos(r)^{\frac{3}{2}+j_L}\sin(r)^{-(\frac{3}{2}+j_L)}\sin(2 r)^{-j_L-\frac{1}{2}}
-\frac{1}{2}\cos(r)\sin(r)A_4^{(j_L,-j_L,-q_R)}(r).\nonumber\\
\eea
The last equation is singular at the North or South Poles $r=0,\pi$ which implies that $A^0_3=0$. So
\bea
A_3^{(j_L,-j_L,-q_R)}(r)&=&-\frac{1}{2}\cos(r)\sin(r)A_4^{(j_L,-j_L,-q_R)}(r).
\eea
Next $\mathcal{E}_3=0$ and $\mathcal{E}_4$ can be solved to give
\bea
A_4^{(j_L,-j_L,-q_R)}(r)=( a_4^0\cos(r)-b_4^0\, _2F_1\left(-\frac{1}{2},-2 \text{j}_L;\frac{1}{2};\cos
   ^2(r)\right))\sin(r)^{-3-2j_L}.
\eea
$a_4^0$ comes with $\cos(r)$ while $b_4^0$ is multiplied by a polynomial in $\cos(r)^2$, this means that the absence of singularity at $r=0$ or $r=\pi$ implies that $a^0_4=0,b^0_4=0$.\\
Similarly for $q_L=-j_L$ we get a singular solutions. So far there is no Kernel, now we go to $|q_L|<j_L$ case.\\
iii)  $|q_L|<j_L$\\
For the isolated case of $q_L=0$ , solving the kernel equations we get the result that the  solution set is empty. The details are given in appendix \ref{appendix:ode4}. 
 Hence the conclusion is that Kernel of $D^{vec}_{10}$ is $\bf{empty}$.\\

\subsection{Cokernel equations}
The Fourier series expansion of  fields contributing to the cokernel is given in eq. (\ref{eq:FourierCok1}), except the follwoing field redefinition
\bea
\bar{c}(\theta,r)=c_{a_{1}}(\theta,r)+2 i q_L c(\theta,r),
\eea
with $c_{a_{1}}(\theta,r)$ an auxiliary variable. 
Now we show that all these equations can be expressed in terms of $SU(2)_L$ generators.
\bea
J^{-}\chi_+&=&e^{- i q_L \psi}e^{- i q_R \phi} l^{-\mu}\partial_{\mu}(e^{ i( q_L+1) \psi}e^{ i q_R \phi}\chi_+(\theta,r)),\nonumber\\
J^{+}\chi_-&=&e^{- i q_L \psi}e^{- i q_R \phi} l^{+\mu}\partial_{\mu}(e^{ i( q_L-1) \psi}e^{ i q_R \phi}\chi_-(\theta,r))
\nonumber\\
J^{-}\chi_3&=&e^{- i (q_L-1) \psi}e^{- i q_R \phi} l^{-\mu}\partial_{\mu}(e^{ i( q_L) \psi}
e^{ i q_R\phi}\chi_3(\theta,r)),\nonumber\\
J^{+}\chi_3&=&e^{- i (q_L+1) \psi}e^{- i q_R \phi} l^{+\mu}\partial_{\mu}(e^{ i( q_L) \psi}
e^{ i q_R\phi}\chi_3(\theta,r))\nonumber\\
J^{-}c_{a_{1}}&=&e^{- i (q_L-1) \psi}e^{- i q_R \phi} l^{-\mu}\partial_{\mu}(e^{ i( q_L) \psi}
e^{ i q_R\phi}c_{a_{1}}(\theta,r)),\nonumber\\
J^{+}c_{a_{1}}&=&e^{- i (q_L+1) \psi}e^{- i q_R \phi} l^{+\mu}\partial_{\mu}(e^{ i( q_L) \psi}
e^{ i q_R\phi}c_{a_{1}}(\theta,r))\nonumber\\
J^{+}J^{-}\text{c}&=&e^{- i (q_L) \psi}e^{- i q_R \phi} l^{+\mu}D_{\nu}( l^{-\nu}D_{\mu}(e^{ i (q_L)\psi}  e^{ i (q_R)\phi}\text{c}(\theta,r))),\nonumber\\
J^{-}J^{+}\text{c}&=&e^{- i (q_L) \psi}e^{- i q_R \phi} l^{-\mu}D_{\nu}( l^{+\nu}D_{\mu}(e^{ i (q_L)\psi}  e^{ i (q_R)\phi}\text{c}(\theta,r))),\nonumber\\
J^3J^3\text{c}&=&q_L^2\text{c}(\theta,r).\nonumber\\
\eea
Next the Cokernel equations can be written in terms of these generators as
\bea
C\mathcal{E}_1&=&-\frac{1}{8}\sin(r)\sin(\theta)(-2 J^-(c_{a_{1}})+2 i J^-(\chi_3)+ i (-2(-1+q_L)\cos(r)\chi_-(\theta,r)\nonumber\\
&+&\sin(r)\partial_r\chi_-(\theta,r))),\nonumber\\
C\mathcal{E}_2&=&-\frac{1}{8}\sin(r)\sin(\theta)(-2 J^+(c_{a_{1}})-2 i J^+(\chi_3)+ i (2(1+q_L)\cos(r)\chi_+(\theta,r)\nonumber\\
&+&\sin(r)\partial_r\chi_+(\theta,r))),\nonumber\\
C\mathcal{E}_3&=&-\frac{1}{4}\tan(r)\sin(\theta)(-i J^-(\chi_+)+ i J^+(\chi_-)-  (2 q_L\cos(r)c_{a_{1}}(\theta,r)\nonumber\\
&-&i\sin(r)\partial_r\chi_3(\theta,r))),\nonumber\\
C\mathcal{E}_4&=&-\frac{1}{8}\sin(r)^2\sin(\theta)(-i J^-(\chi_+)- i J^+(\chi_-)+  (-2 i q_L\cos(r)\chi_3(\theta,r)\nonumber\\
&+&\sin(r)\partial_rc_{a_{1}}(\theta,r))),\nonumber\\
C\mathcal{E}_5&=&-\frac{1}{8}\sin(r)^3\sin(\theta)(\frac{4}{\sin(r)^2}(\frac{1}{2}(J^{+}J^{-}\text{c}+J^{-}J^{+}\text{c})\nonumber\\
&+&J^3J^3\text{c})+(-2  i q_L \text{c}_a(\theta,r)+3\chi_3(\theta,r)-3\cot(r)\partial_r\text{c}(\theta,r)\nonumber\\
&-&\partial^2_r\text{c}(\theta,r))).
\eea

We can express the fields in terms of scalar harmonics$ (i.e. j_R=j_L)$ whose $\psi$ and $\phi$  coordinates dependence has already been extracted above. Similar to the case of kernel equations,
\bea
\chi_+(\theta,r)&=&Y^{(j_L,q_L+1,q_R)}(\theta)\chi_+^{(j_L,q_L+1,q_R)}(r),\quad
\chi_-(\theta,r)=Y^{(j_L,q_L-1,q_R)}(\theta)\chi_-^{(j_L,q_L-1,q_R)}(r),\nonumber\\
\chi_3(\theta,r)&=&Y^{(j_L,q_L,q_R)}(\theta)\chi_3^{(j_L,q_L,q_R)}(r),\quad\qquad
c_{a_{1}}(\theta,r)=Y^{(j_L,q_L,q_R)}(\theta)c_{a_{1}}^{(j_L,q_L,q_R)}(r),\nonumber\\
\text{c}(\theta,r)&=&Y^{(j_L,q_L,q_R)}(\theta)\text{c}^{(j_L,q_L,q_R)}(r).\nonumber\\
\eea
using the inventory of various identities  given in appendix \ref{identities}. Let's begin the analysis.\\
$\alpha$) $q_L=j_L+1$\\
In this case $C\mathcal{E}_1=0,C\mathcal{E}_2=0,C\mathcal{E}_4=0,C\mathcal{E}_5=0$ give empty solution set and $C\mathcal{E}_3=0$  can be solved to give
the follwoing solution for $\chi_-^{(j_L,j_L,q_R)}(r)$
\bea\label{eq:Coksol21}
\chi_-^{(j_L,j_L,q_R)}(r)= \chi_-^0 \sin(r)^{2 j_L}.
\eea
Since the $\hat{Q}^2$ eigenvalue on $\chi_-^{(j_L,j_L,q_R)}(r)$ is $n=2(j_L+1)$, the multiplicity of this solution is
$2j_L+1=n-1$.\\
Siimilarly for $q_L=-j_L-1$ only $\chi_+$ survives and is  given as
\bea\label{eq:Coksol22}
\chi_+^{(j_L,-j_L,q_R)}(r)= \chi_+^0 \sin(r)^{2 j_L},
\eea
With $\hat{Q}^2$ eigenvalue on $\chi_+^{(j_L,-j_L,q_R)}(r)$ as $n=-2j_L-2$, the multiplicity of this solution is $2 j_L+1=|n|-1$.\\
$\beta$) $q_L=j_L$\\
For this value of $q_L$, $C\mathcal{E}_1=0$  can be solved to give
\bea
\chi_-^{(j_L,-1+j_L,q_R)}(r)=-i \cos(r)c_{a_{1}}^{(j_L,j_L,q_R)}(r)-\frac{\sin(r)\partial_r\chi_3^{(j_L,j_L,q_R)}(r)}{2 j_L}.
\eea
Using this value of $\chi_-^{(j_L,-1+j_L,q_R)}(r)$, $C\mathcal{E}_2=0$  yields
\bea
\chi_3^{(j_L,j_L,q_R)}(r)=\chi_3^0\sin(r)^{2 j_L}+i c_{a_{1}}^{(j_L,j_L,q_R)}(r).
\eea
By considering the regularity of the solution at $r=0$ and $r=\pi$ , solving$C\mathcal{E}_3=0$ yields for $c_{a_{1}}^{(j_L,j_L,q_R)}(r)$
\bea
c_{a_{1}}^{(j_L,j_L,q_R)}(r)=\frac{i}{2}\chi_3^0\sin(r)^{2 j_L}+c_{a_{2}}^{(j_L,j_L,q_R)}(r),
\eea
where $c_{a_{2}}^{(j_L,j_L,q_R)}(r)$ is a new function to be determined. Plugging this solution into $C\mathcal{E}_3$, it is converted to a differential equation for $c_{a_{2}}^{(j_L,j_L,q_R)}(r)$.
\bea
&\frac{1}{16 j_L}\sin(r)\sin(\theta)Y^{(j_L,-1+j_L,q_R)}(\theta)(-j_L(7+2 j_L+(-3+2 j_L)\cos(2 r))\nonumber\\
&\times c_{a_{2}}^{(j_L,j_L,q_R)}(r)+\sin(r)(3\cos(r)\partial_rc_{a_{2}}^{(j_L,j_L,q_R)}(r)+\sin(r)\partial^2_rc_{a_{2}}^{(j_L,j_L,q_R)}(r)))=0.\nonumber\\
\eea
We multiply this equation by $c_{a_{2}}^{(j_L,j_L,q_R)}(r)$ and integrate over $r$ from $0$ to $\pi$, if there is smooth solution the result must be zero. On the other hand by partial integrating the term containing $\partial^2_rc_{a_{2}}^{(j_L,j_L,q_R)}(r)$ we get
\bea
-\sin(r)(2 j_L( (2 j_L-3) \cos ^2(r)+5)c_{a_{2}}^{(j_L,j_L,q_R)}(r)^2+\sin(r)^2\partial_rc_{a_{2}}^{(j_L,j_L,q_R)}(r)^2).\nonumber\\
\eea
Note that $2 j_L( (2 j_L-3) \cos ^2(r)+5)$ is always positive for all $j_L$ and all $r$, and $\sin(r)$ is positive, so for this integral to be zero, $c_{a_{2}}^{(j_L,j_L,q_R)}(r)$ must be zero.
\bea
c_{a_{2}}^{(j_L,j_L,q_R)}(r)=0.
\eea
$C\mathcal{E}_4$ is just the complex conjugate of $C\mathcal{E}_3$. so the same analysis goes through. The $C\mathcal{E}_5$ is the conjugate equation to the kernel equation for $\phi_2$ or $\phi_p$ and is given for $q_L=j_L$ by
\bea
&C\mathcal{E}_5=\frac{1}{16}\sin(r)\sin(\theta)Y^{(j_L,j_L,q_R)}(\theta)(8 j_L(1+j_L)\text{c}^{(j_L,j_L,q_R)}(r)
+\sin(r)\nonumber\\
&\times(-6\cos(r)\partial_r\text{c}^{(j_L,j_L,q_R)}(r)+\sin(r)((3+2 j_L)\chi_3^0\sin(R)^{2j_L}-
2 \partial^2_r\text{c}^{(j_L,j_L,q_R)}(r).\nonumber\\
\eea
If the homogeneous piece is zero then this is just the Laplacian, which has not smooth solution on $S^4$. So it is sufficient to construct one smooth solution of the inhomogeneous  equation and that will be the unique solution. It is given by
\bea
\text{c}^{(j_L,j_L,q_R)}(r)=-\frac{1}{4 j_L}\chi_3^0\sin(r)^{2 j_L}.
\eea
Summarizing the  solution set for $q_L=j_L$ is
\bea\label{eq:Coksol1}
\chi_-^{(j_L,j_L-1,q_R)}(r)&=&0,\quad
\chi_3^{(j_L,j_L,q_R)}(r)=\frac{1}{2}\chi_3^0\sin(r)^{2 j_L},\nonumber\\
\text{c}^{(j_L,j_L,q_R)}(r)&=&-\frac{\chi_3^0\sin(r)^{2 j_L}}{4 j_L}.
\eea
For $q_L=-j_L$ \\
We get only one new solution from this 
\bea
\chi_+^{(j_L,-j_L-1,q_R)}(r)&=&0.
\eea
In this case the $Q^2$ eigenvalue$=n=2 j_L$ and the multiplicity is $|n|+1$. \\
Combining the results for $q_L=\pm(j_L+1)$ and $q_L=\pm(j_L)$ the total multiplicity for $\hat{Q}^2$ is $(|n|-1)+(|n|+1)=2|n|.$\\
$\gamma$) $|q_L|<j_L$\\
For $q_L=0$ there is no solution of the cokernel equations. Next we consider the case when $q_L\ne 0$ by following the same argument as given in \ref{appendix:ode4} one can show that solution set is empty for this range of $q_L$.
For round $S^4$ solution set of  only cokernel differential equations is non-empty. \\ Eq.(\ref{eq:Coksol1}) shows that for $qL=\pm j_L$ the solution set for $\chi_3,c,\bar{c}$ depends on a single constant parameter and we will count it only once.  For our choice of normalization the eigenvalue of  $Q^2=n=\pm2 j_L$, with multiplicity $2j_L+1=|n|+1$. Therefore it will contribute a factor $(n+i \alpha.a)^{n+1}$. Similarly  from eqs.(\ref{eq:Coksol21}),(\ref{eq:Coksol22}), for $q_L=\pm(j_L+1)$ we have eigenvalue $Q^2=n=\pm2(j_L+1)$, with multiplicity $2j_L-1=|n|-1$ and the corresponding contribution  $(n+i \alpha.a)^{n-1}$. Using this data the one loop determinant for round $S^4$ can be immediately written down
\bea
Z^{vec-1-loop}_{roundS^4}&=&\prod_{\alpha\in \Delta} \prod_{n\ne 0}(n+i\alpha.a)^{|n|+1}(n+i\alpha.a)^{|n|-1}\nonumber\\
&=& \prod_{\alpha\in \Delta} \prod_{n\ne 0}(n+i\alpha.a)^{2|n|}.
\eea
where $\Delta$ is the set of roots of $G$, which matches with Pestun's result \cite{Pestun:2007rz}.

\section{Hypermultiplet contribution}\label{matteranalysis}
For matter multiplet  the fields in the kernel and cokernel  of $D^{hyper}_{10}$ in cohomological form are
\bea
q_{IA}=\left(
\begin{array}{cc}
q_{11}(\psi ,\theta ,\phi ,r) & q_{12}(\psi ,\theta ,\phi ,r) \\
 q_{21}(\psi ,\theta ,\phi ,r) & q_{22}(\psi ,\theta ,\phi ,r) \\
\end{array}
\right),
\Sigma_{IA}=\left(
\begin{array}{cc}
\Sigma_{11}(\psi ,\theta ,\phi ,r) & \Sigma_{12}(\psi ,\theta ,\phi ,r) \\
 \Sigma_{21}(\psi ,\theta ,\phi ,r) & \Sigma_{22}(\psi ,\theta ,\phi ,r) \\
\end{array}
\right).
\eea
The $\hat{\textbf{Q}}^2$ action is given by:
 \bea\label{eq:q2h}
\hat{\textbf{Q}}^2 q&=&2 e^{-i \tau_3 \psi/2}\partial_{\psi}( e^{i \tau_3 \psi/2} q),\nonumber\\
\hat{\textbf{Q}}^2\Sigma&=&2 e^{-i \tau_3 \psi/2}\partial_{\psi} (e^{i \tau_3 \psi/2} \Sigma).
\eea
Notice here too the shift due to the R-charge. 
The relevant kernel and cokernel equations are obtained by varying the localizing fermionic field of eq. (
\ref{eq:locusmatter}) with
respect to $\Sigma$'s and $q$'s respectively\footnote{ Explicit expression for $V^{hyper}$ is given in appendix \ref{appendix:fermfunctional} }.
\subsection{Analysis of kernel and cokernel equations}
Since the set of fields with flavor index $I=2$  form a copy of  that of $I=1$ we will discuss the kernel and co-kernel equations for $I=1$ only. After Fourier transforming in coordinates $\phi$ and $\psi$ as follows:
\bea
&q_{11}(\psi ,\theta ,\phi ,r)=\text{q}_{11}(\theta ,r) e^{-i \left(\text{q}_L+\frac{1}{2}\right) \psi -i \text{q}_R \phi },\quad
&q_{21}(\psi ,\theta ,\phi ,r)=\text{q}_{21}(\theta ,r) e^{-i \left(\text{q}_L-\frac{1}{2}\right) \psi -i \text{q}_R \phi }.\nonumber\\
\eea
\bea
&\Sigma_{11}(\psi ,\theta ,\phi ,r)=\Sigma_{11}(\theta ,r) e^{-i \left(\text{q}_L-\frac{1}{2}\right) \psi -i \text{q}_R \phi },\quad 
&\Sigma_{21}(\psi ,\theta ,\phi ,r)=\Sigma_{21}(\theta ,r) e^{-i \left(\text{q}_L+\frac{1}{2}\right) \psi -i \text{q}_R \phi }.\nonumber\\
\eea
the kernel equations become:
\bea\label{eqkernel}
& \frac{1}{32} \sin ^2(r) \left(\text{q}_{11}(\theta ,r) (\cos (\theta )+2 \text{q}_L \cos (\theta )-2 \text{q}_R)+i \sin (\theta ) \left(-2 i \partial_{\theta}\text{q}_{11}(\theta
   ,r)-\sin (r) \partial_r\text{q}_{21}(\theta ,r)\right)\right)\nonumber\\
 &+\left(\left((2 \text{q}_L-1) \cos (r) \text{q}_{21}(\theta ,r)\right)\right)=0,\nonumber\\
  &\frac{1}{32} \sin ^2(r) \left(\text{q}_{21}(\theta ,r) (\cos (\theta )-2 \text{q}_L \cos (\theta )+2 \text{q}_R)-i \sin (\theta ) \left(\sin (r) \partial_r\text{q}_{11}(\theta
   ,r)+(2 \text{q}_L+1) \cos (r) \text{q}_{11}(\theta ,r)\right)\right)\nonumber\\
 &+2 i\left(\left( \partial_{\theta}\text{q}_{21}(\theta ,r)\right)\right)=0.
\eea
and  the cokernel are:
\begin{align}
\begin{split}
&\frac{1}{32} \sin ^2(r) \left(\Sigma_{11}(\theta ,r) (\cos (\theta )-2 q_L \cos (\theta )+2 \text{q}_R)+\sin (\theta ) \left(2 \partial_{\theta}\Sigma_{11}(\theta ,r)-i \sin
   (r) \partial_r\Sigma_{21}(\theta ,r)\right)\right)\nonumber\\
  & +2 i \left(\left((\text{q}_L-1) \cos (r) \Sigma_{21}(\theta ,r)\right)\right)=0,\nonumber\\
&\frac{1}{32} \sin ^2(r) \left(\sin (\theta ) \left(2 \partial_{\theta}\Sigma_{21}(\theta ,r)-i \sin (r) \partial_r\Sigma_{11}(\theta ,r)\right)-2 i (\text{q}_L+1) \sin (\theta ) \cos
   (r) \Sigma_{11}(\theta ,r)\right)\nonumber\\
& +\left(\Sigma_{21}(\theta ,r) (\cos (\theta )+2 \text{q}_L \cos (\theta )-2 \text{q}_R)\right)=0.
\end{split}
\end{align}
As in the vector multiplet case,  the isometry group $SO(4)\simeq SU(2)_L\times S(U(2)_R$ of  foliated $S^3$s plays an important role in solving these equations.
It turns out that kernel and cokernel equations can be written in terms of generators of $SU(2)_L$ , whereas the $SU(2)_R$ remains a spectator. For this reason the degeneracy of the solutions to these equations is determined by $q_R$ quantum number.\\
To convert these partial  differential equations into ordinary ones in the variable $\text{r}$ we further expand the fields in terms of spherical harmonics:
\bea
q_{11}(\theta,r)&=&\text{q}_{11}(r)^{\left(\text{j}_L,-\text{q}_L-\frac{1}{2},-\text{q}_R\right)}(r) Y^{\left(\text{j}_L,-\text{q}_L-\frac{1}{2},-\text{q}_R\right)}(\theta ),\nonumber\\
q_{21}(\theta,r)&=&\text{q}_{21}^{\left(\text{j}_L,-\text{q}_L-\frac{1}{2},-\text{q}_R\right)}(r) Y^{\left(\text{j}_L,-\text{q}_L+\frac{1}{2},-\text{q}_R\right)}(\theta ).
\eea
\bea
\Sigma_{11}(\theta,r)&=&\Sigma_{11}^{\left(\text{j}_L,-\text{q}_L-\frac{1}{2},-\text{q}_R\right)}(r) Y^{\left(\text{j}_L,\text{q}_L-\frac{1}{2},-\text{q}_R\right)}(\theta ),\nonumber\\
\Sigma_{21}(\theta,r)&=&\Sigma_{21}^{\left(\text{j}_L,-\text{q}_L-\frac{1}{2},-\text{q}_R\right)}(r) Y^{\left(\text{j}_L,\text{q}_L+\frac{1}{2},-\text{q}_R\right)}(\theta ).
\eea
and get the solutions which we summarize below. \\For kernel equations, solutions, which are regular at the North or South poles of $S^4$, exists only for $q_L=\pm (j_L+\frac{1}{2})$.\\
For $q_L=j_L+\frac{1}{2}$
\bea
\text{q}_{21}^{(\text{j}_L,-\text{j}_L,-\text{q}_R)}(r)= C_2 \sin ^{2 \text{j}_L}(r),
\eea
with   eigenvalue for the  $\hat{Q}^2$ action equal to $ (2j_L+1)$.\\
For  $q_L=-(j_L+\frac{1}{2})$
\bea
\text{q}_{11}^{(\text{j}_L,\text{j}_L,-\text{q}_R)}(r)= C_3 \sin ^{2 \text{j}_L}(r),\quad
\eea
with  eigenvalue for the  $\hat{Q}^2$ action equal to $-(2 j_L+1)$ for constants $C_1,C_2,C_3,C_4$.\\
Analysis of kernel and cokernel equations with physical boundary at $r=\frac{\pi}{2}$, with regularity at one of the poles, and following the logic of appendix \ref{appendix:ode4} agains shows that only the solution set of kernel  is non-empty.\\
For instance
for  $q_L=(j_L+\frac{1}{2})$: 
\bea
\Sigma_{11}^{(jL, jL, qR)}(r)=s_{11}\sin ^{-2 \text{j}_L-3}(r),
\eea
while for   $q_L=-(j_L+\frac{1}{2})$: 
\bea
\Sigma_{21}^{(jL, jL, qR)}(r)=s_{21}\sin(r) ^{-2 \text{j}_L-3}(r),\quad  
\eea
which are not regular at $r=0$ or $\pi$.
\\
For  $q_L=(j_L-\frac{1}{2})$: we get identical set of ordinary differential equations for two fields
\bea
&\left(\left(2 \text{j}_L^2-\text{j}_L-3\right) \cos (2 r)+2 \text{j}_L^2-3 \text{j}_L+4\right) \Sigma_{21}^{(\text{j}_L,\text{j}_L,\text{q}_R)}(r)-
\sin (r) \left(5 \cos (r)
   \Sigma_{21}^{(\text{j}_L,\text{j}_L,\text{q}_R)'}(r)\right)\nonumber\\
&+ \left( \sin (r) \Sigma_{21}^{(\text{j}_L,\text{j}_L,\text{q}_R)''}(r)\right)=0.\nonumber\\
\eea
with the solution
\bea
    \Sigma_{21}^{(\text{j}_L,\text{j}_L,\text{q}_R)}(r)&=&\frac{c_1 P_{2 \text{j}_L-1}^{\sqrt{4 \text{j}_L^2-4 \text{j}_L+5}}(\cos (r))}{\cos ^2(r)-1}+\frac{c_2 Q_{2 \text{j}_L-1}^{\sqrt{4 \text{j}_L^2-4 \text{j}_L+5}}(\cos
   (r))}{\cos ^2(r)-1}.\nonumber\\
\eea
For  $q_L=-(j_L-\frac{1}{2})$:
\bea
&\left(\left(2 \text{j}_L^2-3 \text{j}_L-2\right) \cos (2 r)+2 \text{j}_L^2-\text{j}_L+3\right)  \Sigma_{21}^{(\text{j}_L,1-\text{j}_L,\text{q}_R)}(r)-\sin (r) \left(5 \cos (r)
   \Sigma_{21}^{(\text{j}_L,1-\text{j}_L,\text{q}_R)'}(r)\right)\nonumber\\
& +\left(\sin (r)  \Sigma_{21}^{(\text{j}_L,1-\text{j}_L,\text{q}_R)''}(r)\right)=0.\nonumber\\
\eea
with the solution
\bea
   \Sigma_{21}^{(\text{j}_L,1-\text{j}_L,\text{q}_R)}(r)&=&\frac{c_1 P_{2 (\text{j}_L-1)}^{\sqrt{4 \text{j}_L^2-4 \text{j}_L+5}}(\cos (r))}{\cos ^2(r)-1}+\frac{c_2 Q_{2 (\text{j}_L-1)}^{\sqrt{4 \text{j}_L^2-4 \text{jL}+5}}(\cos
   (r))}{\cos ^2(r)-1}.\nonumber\\
\eea
where $P$ and $Q$ are Legendre functions. These solutions are not regular at $r=0$ or $\pi$ for the case of round $S^4$ or at one pole and the equator $r=\frac{\pi}{2}$ in case of half-$S^4$. Therefore the solution set is empty for cokernel equations.
\section{Wave function on hemisphere $HS^4$}\label{wave functions}
As is clear from the above analysis that for the round $S^4$ solution set of  kernel equations is empty and that for the cokernel equations is nonempty. In order for the boundary to preserve supersymmetry, the  component of the supercurrent normal to the boundary must vanish. Also the boundary conditions must be consistent with the localization locus given in eqs.(\ref{eq:locusvector}),(\ref{eq:locusmatter}).\\ If we consider hemisphere $HS^4$ with supersymmetric BCs  at $r=\frac{\pi}{2}$, the analysis of  kernel and cokernel equations remains identical except that one has to take account of  possible boundary contributions. Also spectrum  remains same with the result that kernel of $D^{vec}_{10}$ is empty and cokernel solutions set is non-trivial with the same eigenvalues and multiplicities if one imposes supersymmetric boundary conditions. Thus practically the only change is to take the range of coordinate r to be $0\le r\le\frac{\pi}{2}$.
\subsection{Supersymmetric boundary conditions}\label{vectorBC}
\subsection*{Vector multiplet}
First we recall that for manifolds with boundary e.g. $HS^4$ the supersymmetric variation of physical action $\hat{Q}S$ vanishes upto total derivative terms. These total derivative terms break supersymmetry at the boundary unless on adds extra terms to Lagrangian such that the $\hat{Q}$ variation of the modified action vanishes. The other way to get rid of the boundary terms is to impose supersymmetric boundary conditions on all the fields. 
For the vector multiplet the boundary contribution is
\bea
\hat{Q}V^{vec}_{Boundary}&=&-\frac{1}{8} \sin (\theta ) \sin ^2(r) \left(\text{c}(\theta ,r) \left(-6 \cos (r) \Lambda(\theta ,r)+\sin (r) \partial_r\left(\Lambda(\theta ,r)\right)\right)\right)\nonumber\\
&-&\left(\left(\left(2 i
   \text{q}_L \text{A}_r(\theta ,r)\right)\right)+\sin (r) \text{A}_r(\theta ,r) \bar{c}(\theta ,r)+i \text{A}_-(\theta ,r) \chi_+(\theta ,r)+i
   \text{A}_+(\theta ,r) \chi_-(\theta ,r)\right)\nonumber\\
  & -&\left(\sin (r) \partial_r\text{c}(\theta ,r) \Lambda(\theta ,r)+2 \chi_3(\theta ,r) \phi_2(\theta
   ,r)+\cos (r) \chi_3(\theta ,r) \Lambda(\theta ,r)\right).
\eea
where $\Lambda$ is the gauge parameter which takes the scalar zero mode as its value at the localization locus. For supersymmetry consistent BCs the boundary contribution vanishes. Dirichlet type boundary conditions correspond to choosing
\bea
\chi_3(\theta ,\frac{\pi}{2})=0,\quad c(\theta ,\frac{\pi}{2})=0,\quad \bar{c}(\theta ,\frac{\pi}{2})=0,
\eea
while keeping $\chi_+,\chi_-$ arbitrary.
Whereas for Neumann type boundary conditions:
\bea
\chi_+(\theta ,\frac{\pi}{2})=0,\quad \chi_-(\theta ,\frac{\pi}{2})=0,
\eea
keeping $\chi_3,c,\bar{c}$ arbitrary at the boundary.
This can be understood in the following way: \\ at the boundary $r=\frac{\pi}{2}$  the Killing spinor satisfies
\bea\label{eq:xiatboundary}
i \tau^3 \xi_A|_{\frac{\pi}{2}}=\bar{\xi}_A|_{\frac{\pi}{2}},
\eea
which motivates choosing the following BC's on the gaugino 
\bea\label{eq:lambdaatboundary}
i \tau^3 \lambda_A|_{\frac{\pi}{2}}=\pm\bar{\lambda}_A|_{\frac{\pi}{2}}.
\eea
The above conditions on $\chi$ follow from its definition in terms of $\lambda$. Moreover for the consistency of supersymmetry 
 \bea\label{eq:Qlambdaatboundary}
i \tau^3 \hat{Q}\lambda_A|_{\frac{\pi}{2}}=\pm \hat{Q}\bar{\lambda}_A|_{\frac{\pi}{2}},
\eea
it follows  that 
 \bea
 F_{\mu\nu}(\theta,\frac{\pi}{2})=0 \quad \mu,\nu=\psi,\theta,\phi,\qquad \Lambda(\theta,\frac{\pi}{2})=a,\quad \partial_r\phi_2(\theta,\frac{\pi}{2})=0,
 \eea
for the lower sign.
On the other hand, for the upper sign choice,
 \bea
i \tau^3 \lambda_A|_{\frac{\pi}{2}}=\bar{\lambda}_A|_{\frac{\pi}{2}},
\eea
we get:
 \bea
 F_{r\nu}(\theta,\frac{\pi}{2})=0 \quad \mu,\nu=\psi,\theta,\phi,\qquad \phi_2(\theta,\frac{\pi}{2})=a,\quad \partial_r\Lambda(\theta,\frac{\pi}{2})=0.
 \eea
 So, for lower sign choice we get Dirichlet and for the upper sign we get Neumann.\\
 If we act once more with $\hat{Q}$  they are closed and trivially satisfied. For example acting  on eq.(\ref{eq:Qlambdaatboundary}), 
 \bea\label{eq:Q2lambdaatboundary}
 i \tau^3 \hat{Q}^2\lambda_A|_{\frac{\pi}{2}}&=&\pm \hat{Q}^2\bar{\lambda}_A|_{\frac{\pi}{2}},\nonumber\\
  i \tau^3 \partial_{\psi}\lambda_A|_{\frac{\pi}{2}}&=&\pm \partial_{\psi}\bar{\lambda}_A|_{\frac{\pi}{2}}.
 \eea
The second line holds upto a constant gauge transformation $a$.
we see that it is trivially satisfied for Dirichlet BCs. Similar is the case for Neumann BCs. Therefore these boundary conditions are closed under  the action of supersymmetry and hence consistent with it.  However there is one subtle point about the SUSY closure of BCs. Since we are working in Euclidean signature and the fields entering the Lagrangian are analytically continued for Lorentzian signature, the BCs imposed on fields are closed under SUSY only if we take the fields as complex valued functions. If one tries to impose BCs on real and  imaginary parts of various fields separately, it turns out that they are not closed under SUSY and will generate infinite number of  differential constraints on gauge field and gaugino at the boundary. In other words the BC in eq. (\ref{eq:Qlambdaatboundary}) is written in covariant form and due to this reason it is closed under SUSY trivially as shown in eq.(\ref{eq:Q2lambdaatboundary}). On the other hand if we do not work covariantly and instead consider the action of $Q$ on $\partial_r\phi_2(\theta,\frac{\pi}{2})=0$ it is easy to see that
\bea
\partial_rQ\phi_2(\theta,\frac{\pi}{2})\ne0,
\eea
unless we impose extra BC on $\lambda_A$
\bea
i \tau^3 \partial_r\lambda_A|_{\frac{\pi}{2}}=\mp\partial_r\bar{\lambda}_A|_{\frac{\pi}{2}}.
\eea
Note the important inversion of sign from $\pm$ to $\mp$. Now this BC  should itself  be closed under SUSY. But it is easy to convince oneself that due to the inversion of sign $\mp$ at each step  if we act once more with $Q$ it will generate  BC other than already imposed and infact one has to impose infinite number of boundary conditions. \\
 However it turns out that our solution set of kernel and co-kernel equations  satisfy BCs irrespective of whether we impose them covariantly or and separately in terms of real and imaginary parts of individual fields.\\
Second constraint that the BC conditions have to satisfy is the action principle.
Taking  arbitrary variations $\delta$ of the fields in the the action, for round metric, to get the equations of motion, we obtain boundary contributions coming from integration by parts. Keeping in mind that the boundary conditions have to be consistent with saddle point solutions and that the later break the gauge symmetry G to its maximal torus, some non abelian terms drop out and we get the following
\bea
\delta \mathcal{L}= 2 i \lambda^A \sigma^r\delta\bar{\lambda}_A+2 F^{r\mu}\delta A_{\mu}-4\delta\bar{\phi} \partial_r\phi -4\delta\phi \partial_r\bar{\phi} .
\eea
 With the set of BCs (\ref{eq:xiatboundary}),(\ref{eq:lambdaatboundary}),(\ref{eq:Qlambdaatboundary}), the boundary term from the action principle vanishes. $\delta$ is an arbitrary variation in the sense that it may represent supersymmetry variation $Q$ too.
 SUSY variation of the $\hat{Q}V$ produces total derivative in the $\psi,\theta,$ or $\phi$ direction and not in $r$ direction. Hence there is no non-trivial contribution from here.\\
  \subsection*{Hyper multiplet}
In the case of hyper multiplet for the boundary conditions to preserve $N=2$ SUSY in $3-d$ at  $r=\frac{\pi}{2}$,  one has to impose complementary BC's on scalars with different $R$-charges .i.e. Dirichlet BC's on the scalars $q_{11},q_{21}$ and Neumann BC's on $q_{21},q_{22}$ or vice versa. Only this choice of BC's satisfy the constraints coming from the vanishing of supercurrent  normal to the boundary \cite{Gaiotto:2008sa,Gaiotto:2014gha}.\\
  In either case we get $\frac{((2n+1)+(2n-1))}{2}=2n$ multiplicity of the $\hat{Q}^2$ eigenstates with eigenvalue $n+i a.\rho$, with $\rho$ the wight vector of the complex conjugate representation $\mathcal{R}$ of the Gauge group G. The one loop factor for hypermultiplet on hemisphere with Dirichlet or Neumann boundary conditions at the equator can immediately be written down
\bea
Z^{HS^4}_{hyper}=\bigg(\prod_{\rho\in weights}\frac{1}{H(i a.\rho)}\bigg)^{\frac{1}{2}}.
\eea
When one reads off the cokernel equations from $\hat{Q}V_{matter}$, an integration by parts is done, which in the case of a manifold with boundary produces  following boundary terms
\bea
\hat{Q}V_{matter}^{Boundary}&=&(\frac{1}{32} i \sin (\theta ) \sin ^3(r) (-\text{q}_{11}(\theta ,r) \Sigma_{22}(\theta ,r)+\text{q}_{12}(\theta ,r) \Sigma_{21}(\theta ,r)-\text{q}_{21}(\theta ,r)
   \Sigma_{12}(\theta ,r)\nonumber\\
  &+&\text{q}_{22}(\theta ,r) \Sigma_{11}(\theta ,r)))|_{r=\frac{\pi}{2}}.
\eea
The discussion of $F_A$ auxiliary field  is not important in boundary conditions and we will no more discuss it.
There are two choices for the boundary conditions for which 
\bea
\hat{Q}V_{matter}^{Boundary}=0.
\eea
There are two choices:\\
(1). $\psi_{\alpha I}|_{r=\frac{\pi}{2}}=+i \bar{\psi}_{\alpha J}\tau^{3J}_I|_{r=\frac{\pi}{2}}$ \\
If we act with supersymmetry $Q$ on this BC,
\bea
Q\psi_{\alpha I}|_{r=\frac{\pi}{2}}=-i Q\bar{\psi}_{\alpha J}\tau^{3J}_I|_{r=\frac{\pi}{2}},
\eea
 it will be closed if we choose Dirichlet BCs on the following fields
\bea
\text{q}_{12}(\theta ,\frac{\pi}{2})&=&0,\quad  \text{q}_{21}(\theta ,\frac{\pi}{2})=0,\quad  \Sigma_{11}(\theta ,\frac{\pi}{2})=0\quad  \Sigma_{22}(\theta ,\frac{\pi}{2})=0,\nonumber\\
\partial_{\theta}\text{q}_{12}(\theta ,\frac{\pi}{2})&=&0,\quad \partial_{\theta}\text{q}_{21}(\theta ,\frac{\pi}{2})=0,\quad\partial_{\theta}\Sigma_{11}(\theta ,\frac{\pi}{2})=0,\quad \partial_{\theta}\Sigma_{22}(\theta ,\frac{\pi}{2})=0,
\eea
and Neumann BCs on the following
\bea
\partial_r\text{q}_{11}(\theta ,\frac{\pi}{2})=0,\quad \partial_r\text{q}_{22}(\theta ,\frac{\pi}{2}),\quad \partial_r\Sigma_{12}(\theta ,\frac{\pi}{2})=0,\quad \partial_r\Sigma_{21}(\theta ,\frac{\pi}{2})=0.
\eea
Acting once more with a supersymmetry operator $Q$ we get
\bea\label{eq:Q2consistency}
Q^2\psi_{\alpha I}|_{r=\frac{\pi}{2}}=-i Q^2\bar{\psi}_{\alpha J}\tau^{3J}_I|_{r=\frac{\pi}{2}}.
\eea
However note that 
\bea
Q^2\psi_{IA}=2\partial_{\psi}\psi_{\alpha A}-i\psi_{IA},\quad Q^2\bar{\psi}_{\alpha A}=2\partial_{\psi}\bar{\psi}_{IA}+i\bar{\psi}_{IA},
\eea
which means that  eq.(\ref{eq:Q2consistency}) is automatically satisfied. However there  is a caveat here. Note that if we act with $Q$ on say
\bea
\partial_r Qq_{11}(\theta,\frac{\pi}{2})|_{r=\frac{\pi}{2}}=\partial_r\Sigma_{11}(\theta,\frac{\pi}{2})\ne 0,
\eea
using the BCs in case $(1)$. It is clear that this will go on to produce infinite number of boundary conditions, not closed within themselves.\\
The resolution is that in working on $HS^4$ with Euclidean signature, the real and imaginary parts of all fields on the Lorentzian signature get mixed when they are analytically continued to Euclidean signature. So when we check the SUSY closure of BCs on the individual fields thinking of them the same way as on Lorentzian space-time, the SUSY fails to close.
On the other hand the full  covariant expression for BCs $\psi_{\alpha I}|_{r=\frac{\pi}{2}}=+i \bar{\psi}_{\alpha J}\tau^{3J}_I|_{r=\frac{\pi}{2}}$ is closed under SUSY by construction. So the conclusion is that the BCs can closed under supersymmetry when written in covariant form in the Euclidean signature.
\\\\
(2) $\psi_{\alpha I}|_{r=\frac{\pi}{2}}=-i \bar{\psi}_{\alpha J}\tau^{3J}_I|_{r=\frac{\pi}{2}}$\\
If we act with supersymmetry $Q$ on this BC, it will be closed if we choose Dirichlet BCs on the following fields
\bea
\text{q}_{11}(\theta ,\frac{\pi}{2})&=&0,\quad  \text{q}_{22}(\theta ,\frac{\pi}{2})=0,\quad  \Sigma_{12}(\theta ,\frac{\pi}{2})=0\quad  \Sigma_{21}(\theta ,\frac{\pi}{2})=0,\nonumber\\
\partial_{\theta}\text{q}_{11}(\theta ,\frac{\pi}{2})&=&0,\quad \partial_{\theta}\text{q}_{22}(\theta ,\frac{\pi}{2})=0,\quad\partial_{\theta}\Sigma_{12}(\theta ,\frac{\pi}{2})=0,\quad \partial_{\theta}\Sigma_{21}(\theta ,\frac{\pi}{2})=0,
\eea
and Neumann BCs on the following
\bea
\partial_r\text{q}_{12}(\theta ,\frac{\pi}{2})=0,\quad \partial_r\text{q}_{22}(\theta ,\frac{\pi}{2}),\quad \partial_r\Sigma_{11}(\theta ,\frac{\pi}{2})=0,\quad \partial_r\Sigma_{22}(\theta ,\frac{\pi}{2})=0.
\eea
Like that previous case $(1)$ this choice of BCs is closed under supersymmetry except for the properly writing the BCs in a covariant way with respect to Euclidean signature. However fortunately the solution set of kernel and co-kernel equations that we have found satisfy BCs irrespective of whether we impose them covariantly or and separately in terms of real and imaginary parts. 
Applying variational principle to  $S_{hyper}$, to get equations of motion ,we get boundary terms
\bea
\delta\mathcal{L}_{hyper}=\delta q^A D^r q_A-\frac{i}{2}\bar{\psi}\bar{\sigma}^r\delta \psi.
\eea
Choosing one of the above BCs, the action principle will be satisfied as well as these BCs are consistent with supersymmetry.

 \subsection{$Z^{Dir}_{hemi-S^4}$}
Knowing that the eigenvalue of $\hat{Q}^2$ for $\chi_+(\psi,\theta,\phi,r)$ is $n+i a.\alpha$, with corresponding multiplicity $|n|-1$ with $n\in Z$, $a$ the zero mode of the  imaginary part of the scalar of the vector multiplet and $\alpha$ the roots of the gauge group G. The expression for the one loop determinant can be written 
\bea
Z_{vec.-Dir}^{1-loop}&=&\prod_{\alpha\in\Delta}\prod_{n\in \mathbb{Z}_+}(n+i a.\alpha)^{\frac{n-1}{2}}(n-i a.\alpha)^{\frac{n-1}{2}}\nonumber\\
&=& \prod_{\alpha\in\Delta_+}\prod_{n\in \mathbb{Z}_+}(n+i a.\alpha)^{n-1}(n-i a.\alpha)^{n-1}\nonumber\\
&=& \prod_{\alpha\in\Delta_+}\prod_{n\in \mathbb{Z}_+}\frac{(n+i a.\alpha)^{n}(n-i a.\alpha)^{n}}{(n+i a.\alpha)(n-i a.\alpha)}\nonumber\\
\eea
$\Delta$ representing the root system.
The regularized form of this ill defined \cite{Pestun:2007rz}  product is 
\bea
Z_{vec.-Dir}^{1-loop}&=& \prod_{\alpha\in\Delta_+}G(1+i a.\alpha)G(1-i a.\alpha)\Gamma(1+i a.\alpha)\Gamma(1-i a.\alpha).
\eea
Using the identity 
\bea
\frac{1}{\Gamma(1+i a.\alpha)\Gamma(1-i a.\alpha)}=\frac{\sin(i\pi a.\alpha) }{a.\alpha}
\eea
\bea
Z_{vec.-Dir}^{1-loop}&=& \prod_{\alpha\in\Delta_+}G(1+i a.\alpha)G(1-i a.\alpha)\frac{ a.\alpha}{\sin(i \pi a.\alpha)}\nonumber\\
&=& \prod_{\alpha\in\Delta_+} H(i a.\alpha)\frac{a.\alpha}{\sinh( \pi a.\alpha) }
\eea

getting following hemisphere wave function
\bea 
Z^{Dir}_{hemi-S^4}&=&Z_{vec.-Dir}^{1-loop}Z_{inst}\nonumber\\
&=&\prod_{\alpha\in\Delta_+} e^{-\frac{4\pi^2 \tr{a^2}}{g_{YM}^2}}H(i a.\alpha)\frac{a.\alpha}{\sinh( \pi a.\alpha) } 
Z_{inst}^k(a,\tau)
\eea
with $\tau=\frac{\theta}{2\pi}+\frac{4\pi i}{g_{YM}^2}$ and 
$Z_{inst}^k$ is the  contribution of  
$k-th$  sector of the  Nekrasov instanton partition function.\\
Recall that the instanton configurations contributing to the path integral are point-like
instantons localised at the pole of the hemisphere  and in particular  are  pure (large) gauge at the boundary $S^3$.
Given the large gauge transformation $\mathcal{T}$ ,  which maps  the boundary  
$S^3$ to  the $SU(N)$ Lie group,  the corresponding winding number is given as:
\bea
k=\frac{1}{2 \pi^2}\int_{S^3} \tr(\mathcal{T}d\mathcal{T})^3.
\eea



\subsection{$Z^{Neu}_{hemi-S^4}$}\label{NeumannZ}
Neumann BCs by definition imply that the  components of fields tangential to the boundary $r=\frac{\pi}{2}$  are kept arbitrary and consequently in performing the path integral one has to integrate over all field configurations. However to be able to apply localization, the field configurations satisfying some  BCs must be solution of the saddle point equations. As is evident from solution of saddle equations (\ref{eq:locusvector}),(\ref{eq:locusmatter}), the  infinite  dimensional field space is reduced to a single  scalar zero mode $a$. For Dirichlet BCs this zero mode  is fixed  but for Neumann BCs $a$ takes arbitrary values at the boundary and so one has to integrate over it to get the wave function. In general the Neumann wave function depends on the variables canonically conjugate to those fixed by the Dirichlet BCs. If at the boundary we see $4d$ vectormultiplet as composed of  one $3d$ vectormultiplet plus a $3d$ chiral multiplet, then  the Neumann BCs data corresponds to the fixed value of  $3d$ chiral multiplet at the boundary. The dynamical fields of $3d$ chiral multiplet are given in terms of $4d$ fields as
\bea
\{D^i\phi_2,D^r\phi_1,F^{ir},fermionic\quad super-partners\}
\eea 
with $i=\psi,\theta$ and $\phi$. Therefore

\bea
Z_{vec.-Neu}^{1-loop}&=& \prod_{\alpha\in\Delta}\prod_{n\in \mathbb{Z}_+}(n+i a.\alpha)^{\frac{n+1}{2}}(n-i a.\alpha)^{\frac{n+1}{2}}\nonumber\\
&=&  \prod_{\alpha\in\Delta_+}\prod_{n\in \mathbb{Z}_+}(n+i a.\alpha)^{n+1}(n-i a.\alpha)^{n+1}\nonumber\\
&=&  \prod_{\alpha\in\Delta_+}\prod_{n\in \mathbb{Z}_+}(n+i a.\alpha)^{n}(n-i a.\alpha)^{n}(n+i a.\alpha)(n-i a.\alpha)\nonumber\\
\eea

$\Delta$ and $\mathfrak{g}$ representing the root system and Lie algebra respectively, of $SU(N)$.
The  regularized form of this  infinite product is \cite{Pestun:2007rz} 
\bea
Z_{vec.-Neu}^{1-loop}&=&  \prod_{\alpha\in\Delta_+}\frac{G(1+i a.\alpha)G(1-i a.\alpha)}{\Gamma(1+i a.\alpha)\Gamma(1-i a.\alpha)}
\eea

Using the identity 
\bea
\frac{1}{\Gamma(1+i a.\alpha)\Gamma(1-i a.\alpha)}=\frac{\sin(i\pi a.\alpha) }{a.\alpha}
\eea

\bea\label{eq:Neu1}
Z_{vec.-Neu}^{1-loop}&=&  \prod_{\alpha\in\Delta_+}G(1+i a.\alpha)G(1-i a.\alpha)\frac{\sinh(\pi a.\alpha)}{ a.\alpha}\nonumber\\
&=&  \prod_{\alpha\in\Delta_+} H(i a.\alpha)\frac{\sinh(\pi a.\alpha) }{a.\alpha}
\eea

For Neumann BC's, instanton configurations contributing to the path integral are again localised at the pole
and are pure (large gauge)  at the equator  $S^3$, at  $r=\frac{\pi}{2}$, with winding number $k$ as before.

 Therefore the full partition function is:
\bea
Z^{Neu}_{hemi-S^4}=\int_{\mathfrak{g}} da  \prod_{\alpha\in\Delta_+}e^{-\frac{4\pi^2 \tr{a^2}}{g_{YM}^2}} H(i a.\alpha)\frac{\sinh( \pi a.\alpha) }{a.\alpha}Z_{inst}(a,\tau)
\eea

and where  $Z_{inst}$ is the full holomorphic part of  Nekrasov partition function.  Note that here we have written the Neumann wave function after summing over all instanton sectors and thus getting $Z_{int}(a,\tau)$. However in principle   Neumann wave function is computed for each instanton sector labelled by an integer $k$ and hence depends on the discrete parameter $k$. After summing over all values of $k$ one gets the last expression.



\subsection{Large Radius limit $R\to\infty$}
In our computation we have set the radius $R_{S^4}\equiv R=\frac{1}{\epsilon}=1$.  For illustration let's take $G=SU(2)$ in this subsection. Then restoring  it one gets the following expressions for the bulk one-loop part of $HS^4$ with Dirichlet and Neumann BCs.
\bea 
Z^{Dir}_{1-loop}
&=& H(i 2 R a)\frac{2 \pi a R}{\sinh(2 \pi R a) }
 ,\nonumber\\
Z^{Neu}_{1-loop}&=&  H(i 2 R a)\frac{\sinh( 2 \pi R a) }{2 \pi R a}.
\eea
Now using the following identities
\bea
\ln H(x)&=&-x^2\ln(|x|)e^{\gamma-\frac{1}{2}}+\mathcal{O}(\ln(|x|))\quad x\to\infty,\nonumber\\
\ln (\frac{2 \pi x}{\sinh(2 \pi x) })^{\pm}&=&\pm\ln (2\pi x)\mp2\pi x \quad x\to\infty.\nonumber\\
\eea
it is easy to see that to leading order in $R\to\infty$ limit we get the following simplified expressions
\bea 
Z^{Dir}_{1-loop}=e^{2 a^2 R^2\ln (2 |a| R) e^{\gamma-\frac{1}{2}}},\nonumber\\
Z^{Neum}_{1-loop}=e^{2 a^2 R^2\ln (2 |a| R) e^{\gamma-\frac{1}{2}}}.
\eea
where the positive sign in the exponential is accounted by taking $a$ to be anti-hermitian.\\
So we reach an interesting conclusion that  at leading order in $R\to\infty$ limit Dirichlet and Neumann BCs lead to same perturbative result. This exponential contribution of one loop in the large radius limit can be interpreted as producing an RG flow that  renormalizes the classical gauge  coupling constant $g_{YM}$.
\section{Gluing back  two hemisphere $HS^4$ wave functions}\label{sec:joineds4}
\subsection*{Wave functions with Dirichlet BCs}
If we see $N=2,d=4$ vector multiplet as a combination of an $N=2,d=3$ vector and a chiral  multiplet, then imposing Dirichlet BC's at the equator $r=\frac{\pi}{2}$ of $S^4$ amounts to freezing the $3d$ vector multiplet to fixed value and consequently decoupling the gauge theory dynamics from two sides of the equator. Gluing the two wave functions then naturally implies that one has to put back $N=2,d=3$ gauge multiplet at the equator. See also the discussion in \cite{Cabo-Bizet:2016ars}.\\
It has being argued that the two wave functions for Dirichlet BC's are glued along the equator $r=\frac{\pi}{2}$ by gauging the global symmetry under which the boundary values of the dynamical fields transform \cite{Drukker:2010jp,Bullimore:2014nla}. In other words one has to put an $N=2$, $3d$ vector multiplet at the equator
and include the corresponding partition function. In fact, our results confirm this general argument.
As for the matter multiplet, the wave functions from two hemispheres are joined together by turning on a super potential coupling at the equator \cite{Gaiotto:2014gha}. However being a Q-exact term, the superpotential does not contribute  to the localization computation.
\bea\label{eq:glued1}
Z&=& \int_{\mathfrak{g}} d a \bigg[\prod_{\alpha\in \Delta_+}\bigg(H(i a.\alpha)\frac{a.\alpha}{ \sinh(\pi a.\alpha)}e^{-\frac{4\pi^2 \tr{a^2}}{g_{YM}^2}}\frac{\sinh(\pi \alpha.a)^2}{(\alpha.a)^2}e^{-\frac{4\pi^2 \tr{a^2}}{g_{YM}^2}}\frac{a.\alpha}{ \sinh(\pi a.\alpha)}H(i a.\alpha)\bigg)\bigg]\nonumber\\
&\times&\bigg[Z^{HS^4}_{hyper}Z^{HS^4}_{hyper}\bigg]|Z_{inst}|^2\nonumber\\
&=& \int_{\mathfrak{g}} d a \bigg[\prod_{\alpha\in \Delta_+}e^{-\frac{8\pi^2 \tr{a^2}}{g_{YM}^2}}H(i a.\alpha)^2\bigg]
\bigg[\prod_{\rho\in weights}\frac{1}{H(i a.\rho)}\bigg]|Z_{inst}(a,q)|^2\nonumber\\
&=&Z^{S^4}_{pestun}
\eea 
where $q=e^{2\pi i \tau}$, 
The last identity can also be interpreted as  factorization of round sphere $S^4$ partition function, though more precisely 
it is a convolution of two Dirichlet wave functions of two hemispheres with non-trivial integral kernel, the latter being due
to a 3D vector multiplet at the equator. 
Perhaps the instanton contribution requires a comment: one would have naively
thought that in glueing  two Dirichlet wave functions one should have matched the $k$-th instanton sector on one side with
the the $-k$-th anti-instanton sector on the other side. This would have produced a function of $|q|$, i.e. no $\Theta$ dependence.
This is not the $S^4$ answer however, which is not diagonal in the instanton number:  to get the $S^4$ result one has actually to sum over all (anti-)instanton
sectors on each hemisphere before glueing. Put it differently,  the identification of fields at the boundary is up to large gauge transformations.

 
 \subsection*{Wave functions with Neumann BCs}
We will  only be sketchy here to describe the gluing of two Neumann wave functions.
Roughly speaking Neumann BCs are canonically conjugate to Dirichlet BCs. In the semiclassical approximation Neumann wave function is related to Dirichlet wave function through Legendre transformation. From section \ref{NeumannZ} we know that the $4d$ vector multiplet when restricted to $3d$ boundary, can be decomposed into a $3d$ vector plus a $3d$ chiral multiplet. In the same vein Dirichlet wave function depends on the value of  scalar of the $3d$ gauge multiplet at the boundary, whereas Neumann depends on the value  of $3d$ chiral multiplet  at the boundary. These boundary conditions are constrained by  localizing equations and for this  reason  $3d$ chiral vev at the boundary is taken to be zero in our case.\\
Compared to the Dirichlet case, where we needed to include a 3d vector multiplet partition function in the glueing procedure, in the Neumann case we are facing 
an over counting problem, i.e. we count twice the contribution of the boundary 3D vector multiplet, since in this case the
corresponding boundary degrees of freedom  from each side are not frozen, as it is clear from eq.(\ref{eq:Neu1}). Therefore in the glueing
procedure we insert a factor:
\bea
 \prod_{\alpha} \frac{\alpha(a)}{\sinh(\pi (\alpha(a)))}
\eea
to remove the redundant  degrees of freedom.
This glueing procedure gives rise to  the round $S^4$ partition function. However  it would be very nice to have a better understanding of how this measure arises from the full path integral. See e.g. \cite{Floch:2015hwo}.
 \subsection*{Wave functions with Dirichlet-Neumann BCs}
If we impose Dirichlet BCs on the vector multiplet fields on one hemisphere with the resulting one-loop part
 \bea
Z_{vec.-Dir}^{1-loop}= \prod_{\alpha\in\Delta_+} H(i a.\alpha)\frac{a.\alpha}{\sinh( \pi a.\alpha) }
\eea
and on the complementary hemisphere we impose Neumann BCs with the following one-loop part
\bea
Z_{vec.-Neu}^{1-loop}= \prod_{\alpha\in\Delta_+} H(-i a.\alpha)\frac{\sinh(\pi a.\alpha) }{a.\alpha}
\eea
it is obvious that if we naively glue these two hemispheres we get
\bea
Z_{vec.-Dir}^{1-loop}Z_{vec.-Neu}^{1-loop}\approx \prod_{\alpha\in\Delta_+} H(i a.\alpha) H(-i a.\alpha)\approx Z^{vec.}_{S^4}
\eea
This shows that no extra measure is needed to glue  $Z^{vec Dir.}_{HS^4}$ with $Z^{vec Nem.}_{HS^4}$ to get $Z^{vec}_{S^4}$. Intuitively the over counting of modes from  the hemisphere with Neumann BCs is compensated by the removal of boundary modes by imposing Dirichlet BCs on the other hemisphere. However a more satisfactory explanation in terms of path integral will be  illuminating. 
\section{One loop determinant and $SO(5)$ harmonics}\label{5harmonics}
One loop determinants can also be computed more directly using the full $SO(5)$ harmonics. Like the case of $SO(4)$ harmonics as given in the first part of work, we are only interested in the spectrum of $\hat{Q}^2$ on the kernel$(D_{10})$ and cokernel$(D_{10})$. The purpose of this and the next section is to compute the net multiplicity of $\hat{Q}^2$ on kernel and cokernel of $D_{10}^{vec}$ for round $S^4$, hemisphere $HS^4$ with Dirichlet BCs , Neumann BCs at the equator. We show that it matches with the results obtained using $SO(4)$.
The task can be simplified by observing how vector and scalar harmonics of  $SO(4)$ irreps. are embedded in $SO(5)$ irreps. Here it is helpful to recall some useful results from Lie group Representation theory \cite{barut1986theory}.\\
Irreducible representations of $SO(2 k+1)$ determined by their highest weights $(n_1,n_2,...,n_k)$ with integer or half-integer entries, when restricted to the subgroup $SO(2k)$, contains all irreps. of the later with highest weights $(p_1,p_2,...,p_k)$ with integer or half-integer entries, satisfying the following constraints
\bea\label{branching}
n_1\ge p_1\ge n_2\ge p_2...\ge n_{k-1}\ge p_{k-1}\ge |n_k|.
\eea
If $n_i$ are integers ( half integers) so are $p_i$. Quadratic Casimir is an important operator for a Lie algebra whose eigenvalues for different irreps. are used to regularize infinite sums using heat kernel technique.  For irreps. of orthogonal group it is given by
\bea
C_2(n_1,n_2,...,n_{k+1})=n.n+2 w.m
\eea
with Euclidean dot product assumed and where the Weyl vector $w$ given by
\bea
w_i=\left\{
                \begin{array}{ll}
                  k-i+1 \quad for \quad SO(2k+2)\\
                k+\frac{1}{2}-i \quad for \quad SO(2k+1)
                \end{array}
              \right.
\eea
 Assume that the weights are given in the basis of Cartan generators $(j_L^3,j_R^3)$ for $SU(2)_L\times SU(2)_R\sim SO(4)\subset SO(5)$, with $SO(4)$ being the isometry group of $S^3$ at constant value of coordinate $r$. Then the irreps. of scalar and vector  $SO(5)$ harmonics   can be constructed by starting with the simple roots
\bea
(1,0),\qquad (-\frac{1}{2},\frac{1}{2})
\eea
in the above basis. Here we describe only the final results of the construction. First of all it is easy to check that   $SO(5)$  Lie algebra is generated by the following generators
\bea
j_R^+&=&(-e^{i \phi } \csc (\theta ),-i e^{i \phi },e^{i \phi } \cot (\theta ),0),\quad j_R^-=(-e^{-i \phi } \csc (\theta ),i e^{-i \phi },e^{-i \phi } \cot (\theta ),0),\nonumber\\ j_L^+&=&(e^{i \psi } \cot (\theta ),-i e^{i \psi },-e^{i \psi } \csc (\theta ),0),\quad j_L^-=(e^{-i \psi } \cot (\theta ),i e^{-i \psi },-e^{-i \psi } \csc (\theta ),0),\nonumber\\  j_R^3&=&(0,0,1,0),\quad  j_L^3=(1,0,0,0),\nonumber\\
j_5^+&=&e^{\frac{1}{2} i (\phi -\psi )}(\csc \left(\frac{\theta }{2}\right) (-\cot (r)),-2 i \cos \left(\frac{\theta }{2}\right) \cot (r),\csc \left(\frac{\theta }{2}\right) \cot (r),-i \sin \left(\frac{\theta
   }{2}\right)),\nonumber\\
 j_5^-&=&e^{-\frac{1}{2} i (\phi -\psi )}(\csc \left(\frac{\theta }{2}\right) (-\cot (r)),2 i \cos \left(\frac{\theta }{2}\right) \cot (r),\csc \left(\frac{\theta }{2}\right) \cot (r),i \sin \left(\frac{\theta
   }{2}\right)),\nonumber\\
 j_6^+&=&e^{\frac{1}{2} i (\psi +\phi )}(\sec \left(\frac{\theta }{2}\right) (-\cot (r)),-2 i \sin \left(\frac{\theta }{2}\right) \cot (r),\sec \left(\frac{\theta }{2}\right) (-\cot (r)),i \cos \left(\frac{\theta
   }{2}\right)),\nonumber\\
  j_6^-&=&e^{-\frac{1}{2} i (\psi +\phi )}(\sec \left(\frac{\theta }{2}\right) (-\cot (r)),2 i \sin \left(\frac{\theta }{2}\right) \cot (r),\sec \left(\frac{\theta }{2}\right) (-\cot (r)),-i \cos \left(\frac{\theta
   }{2}\right)).
\eea 
\subsection*{$\mathbb{Z}_2$ action}
Keeping in mind the fact that the above generators act on the fields as a differential  and hence the fourth entry corresponds to derivative w.r.t. r, we  conclude  that  first six generators are even under $Z_2$ action $r\to \pi-r$, whereas the last four generators are odd.
\subsection*{Harmonics}
The logic for contstructing $SO(5)$ harmonics is simple. One repeatedly applies negative roots 
$(-\frac{1}{2},-\frac{1}{2}),(\frac{1}{2},-\frac{1}{2})$ to the highest weight state of an $SO(5)$ irrep. to get a state which is a linear combination $SO(4)$ highest weight and $SO(4)$ descendants. One then removes the descendants part to get the irreps. of $SO(4)$ given by its highest weight. This construction has following properties
\begin{itemize}
\item The highest weights of $SO(5)$ both for scalars and vectors are $Z_2$ even.
\item the two lowering operator represented by negative roots $(-\frac{1}{2},-\frac{1}{2}),(\frac{1}{2},-\frac{1}{2})$, which do not belong to $SO(4)$, project out even modes w.r.t. $Z_2$ action.
\item Since we already know from the branching rule given in \ref{branching} which $SO(4)$ irreps. appear in a given $SO(5)$ irrep. one can easily see that by counting how many times one needs to apply these two negative roots to reach an allowed $SO(4)$ highest weight state starting from a given $SO(5)$ highest weight state
\item If the count is even (odd) the corresponding $SO(4)$ irrep. is even(odd).
\end{itemize}
\subsection*{Scalar harmonics}
So for $SO(5)$ case we will only describe the final results. \\
For scalars the $SO(5)$ highest weight state appears only for $j_R=j_L$ and is given by $(j_L,j_L)$
\bea
\cos ^{2 \text{j}_L}\left(\frac{\theta }{2}\right) \sin ^{2 \text{j}_L}(r) e^{i \text{j}_L (\psi +\phi )}
\eea
and is clearly even under $Z_2$ action. So if we apply $j_6^-$ or $j_5^-$  on it the result will be odd. As a result  decomposing this $SO(5)$ representation in terms of irreps. of $SO(4)$ we get the following
\bea
(j_L-\frac{n}{2},j_L-\frac{n}{2})\quad for \quad n=0,1,..,2j_L.
\eea
For n an even integer this irrep. is even under $Z_2$ and for n odd the irrep. is $Z_2$ odd.
\subsection*{Vector harmonics}
Vector $SO(5)$ harmonics come in two classes of $SO(5)$ irreps. labeled by highest weights and they decompose in $SU(2)_L\times SU(2)_R$ irreps. represented by  the highest weights $(j_L,j_R)$, as
\begin{enumerate}
\item First $SO(5)$ irrep. is $(j_L+1,j_L)$, which decomposes into three $SO(4)$ irreps.
\begin{enumerate}
\item $(j_L-\frac{n}{2},j_L+1-\frac{n}{2})$ for $n=0,...,2 j_L$. Irreps. with even n are invariant under $Z_2$, while odd n modes are odd.
\item $(j_L+\frac{1}{2}-\frac{n}{2},j_L+\frac{1}{2}-\frac{n}{2})$ for $n=0,...,2 j_L$. Importantly in this case irreps. with even n are odd and irreps. with odd n are even under $Z_2$ action.
\item $(j_L+1-\frac{n}{2},j_L-\frac{n}{2})$ for $n=0,...,2 j_L$. These are invariant for even n and odd for odd n.
\end{enumerate}
\item Second $SO(5)$ irrep. for vector harmonics is $(j_L,j_L)$ and it decomposes into  $SO(4)$ irreps. as $(j_L-\frac{n}{2},j_L-\frac{n}{2})$ for $n=0,...,2 j_L$, which are even for even n and odd for odd n.
\end{enumerate}
\section{$Z^{vec}_{1-loop}$ via $SO(5)$ harmonics }\label{51loop}
For $SO(4)$ irreps. $(j_L-\frac{n}{2},j_L-\frac{n}{2})\quad for \quad n=0,1,..,2j_L$ contained in $(j_l,j_L)$ irrep. of $SO(5)$,
 various dynamical scalar fields will contribute the following to the net multiplicity of the one-loop determinant. Scalar contribution for  $Z_2$ even irreps. is denoted as $S_e$ and for odd irreps. as $S_o$.
\bea
S_e=\sum_{j=\frac{m}{2}}^{\infty}[(j+1)^2-\frac{m^2}{4}],\quad S_o=\sum_{j=\frac{m}{2}}^{\infty}[\left(j+\frac{1}{2}\right)^2-\frac{m^2}{4}]
\eea
Similarly for  vector harmonics of $SO(5)$ one gets the following individual contribution to the net multiplicity of one loop determinant.\\ For even irreps. of $SO(4)$
\bea
V_e^0=\sum_{j=\frac{m}{2}}^{\infty}[(j+1)^2-\frac{m^2}{4}],\quad V_e^+=\sum_{j=\frac{m}{2}}^{\infty}[(j+2)^2-\frac{(m+2)^2}{4}],\quad V_e^-=\sum_{j=\frac{m}{2}}^{\infty}[(j+1)^2-\frac{(m-2)^2}{4}],\nonumber\\
\eea
and for odd irreps. of $SO(4)$
\bea
V_o^0=\sum_{j=\frac{m}{2}}^{\infty}[(j+\frac{3}{2})^2-\frac{m^2}{4}],\quad V_o^+=\sum_{j=\frac{m}{2}}^{\infty}[(j+\frac{3}{2})^2-\frac{(m+2)^2}{4}],\quad V_e^-=\sum_{j=\frac{m}{2}}^{\infty}[(j+\frac{1}{2})^2-\frac{(m-2)^2}{4}].\nonumber\\
\eea
\subsection*{Regularizing the Infinite sums}
As an example we will describe in detail the regularization of  $S_e$. For the other contributions we will only give the final results. Consider the following expression in the heat kernel regularization
\bea
S_e(t,m)&=&\sum_{j=\frac{m}{2}}^{\infty}[(j+1)^2-\frac{m^2}{4}]e^{-t(2j^2+3 j)}\nonumber\\
&=& \sum_{j=\frac{m}{2}}^{\infty}[(j+\frac{3}{4})^2+\frac{1}{2}(j+\frac{3}{4})+(\frac{1}{16}-\frac{m^2}{4})]e^{-2t(j+\frac{3}{4})^2+\frac{9}{8}t}
\eea
where $2j^2+3j$ is the regularization factor for $j_L=j_R\equiv j$ representation of $SO(5)$.
Taking the Mellin transform of $S_e$ w.r.t. $t$
\bea
\tilde{S}_e(s,m)=\int_0^{\infty}t^{s-1}S_e(t,m)dt
\eea
we get
\bea
\tilde{S}_e(s,m)&=&\sum_{j=\frac{m}{2}}^{\infty}[(j+\frac{3}{4})^2+\frac{1}{2}(j+\frac{3}{4})+(\frac{1}{16}-\frac{m^2}{4})]\Gamma(s)2^{-s}[(j+\frac{3}{4})^2-\frac{9}{16}]^{-s}\nonumber\\
&=& \sum_{j=\frac{m}{2}}^{\infty}\bigg[(j+\frac{3}{4})^{2-2s}+\frac{1}{2}(j+\frac{3}{4})^{1-2s}+(\frac{1}{16}-\frac{m^2}{4})(j+\frac{3}{4})^{-2s}\bigg]\Gamma(s)2^{-s}[1-\frac{9}{16}(j+\frac{3}{4})^{-2}]^{-s}\nonumber\\
\eea
Now applying the Binomial expansion
\bea
[1-\frac{9}{16}(j+\frac{3}{4})^{-2}]^{-s}=\sum_{k}(-1)^k\frac{(-s)!}{k!(-s-k)!}(\frac{9}{16})^k(j+\frac{3}{4})^{-2k}
\eea
and using  $\Gamma$ function analytic continuation
\bea
[1-\frac{9}{16}(j+\frac{3}{4})^{-2}]^{-s}=\sum_{k}(-1)^k\frac{\Gamma(1-s)}{\Gamma(k+1)\Gamma(1-s-k)!}(\frac{9}{16})^k(j+\frac{3}{4})^{-2k}
\eea
Substituting this in the expression for $\tilde{S}$
\bea
\tilde{S}_e(s,m)&=&\sum_{j=\frac{m}{2}}^{\infty}\sum_k\bigg[(j+\frac{3}{4})^{2-2s-2k}+\frac{1}{2}(j+\frac{3}{4})^{1-2s-2k}+(\frac{1}{16}-\frac{m^2}{4})(j+\frac{3}{4})^{-2s-2k}\bigg]\nonumber\\
&\times&\Gamma(s)2^{-s}(-1)^k\frac{\Gamma(1-s)}{\Gamma(k+1)\Gamma(1-s-k)!}(\frac{9}{16})^k
\eea
Since the above summation is absolutely convergent, one can exchange the order of summation and at the same time shift $j$ to $j+\frac{m}{2}$ and perform the $j$ summation in terms of Hurwitz Zeta function to get
\bea
\tilde{S}_e(s,m)&=&\sum_k\bigg[\zeta \left(2 k+2 s-2,\frac{m}{2}+\frac{3}{4}\right)+\frac{1}{2}\zeta \left(2 k+2 s-1,\frac{m}{2}+\frac{3}{4}\right)+(\frac{1}{16}-\frac{m^2}{4})\zeta \left(2 k+2 s,\frac{m}{2}+\frac{3}{4}\right)\bigg]\nonumber\\
&\times&\Gamma(s)2^{-s}(-1)^k\frac{\Gamma(1-s)}{\Gamma(k+1)\Gamma(1-s-k)!}(\frac{9}{16})^k
\eea
In the last step we take the inverse Mellin transform
\bea
S_e(t,m)=\frac{1}{2\pi i}\int_{\mathcal{C}}t^{-s}\tilde{S}_e(s,m)ds
\eea
and perform complex integration along a contour $\mathcal{C}$ which encloses all the poles of integrand. Interestingly when we evaluate the last integral for various poles of $s$, we fine that the series terminates for finite values of $k$. We thus obtain
\bea
S_e(t,m)= \frac{s_{-1}}{t}-\frac{9 \left(16 m^2-13\right)}{2048}\frac{1}{\sqrt{t}}+(\frac{m^3}{12}-\frac{m^2}{16}-\frac{m}{12}+\frac{7}{48})t^0+\frac{1}{64} \ \left(11-8 m^2\right)\sqrt{t}+...
\eea
Following the same procedure we find
\bea
S_o(t,m)=...+(\frac{m^3}{12}+\frac{m^2}{16}-\frac{m}{12}-\frac{7}{48})t^0+...
\eea
and similarly for vector harmonics
\bea
V_e^0&=&...+(\frac{m^3}{12}+\frac{m^2}{16}-\frac{m}{12}-\frac{1}{48})t^0+...,\quad V_e^+=...+(\frac{m^3}{12}+\frac{5 m^2}{16}+\frac{m}{6}-\frac{1}{16})t^0+...,\nonumber\\
V_e^-&=&...+(\frac{m^3}{12}-\frac{7 m^2}{16}+\frac{2 m}{3}-\frac{13}{48})t^0+...
\eea
and 
\bea
V_o^0&=&...+(\frac{m^3}{12}-\frac{m^2}{16}-\frac{m}{12}+\frac{1}{48})t^0+...,\quad V_o^+=...+(\frac{m^3}{12}+\frac{7 m^2}{16}+\frac{2 m}{3}+\frac{13}{48})t^0+...,\nonumber\\
V_o^-&=&...+(\frac{m^3}{12}-\frac{5 m^2}{16}+\frac{m}{6}+\frac{1}{16})t^0+...
\eea
In the application of localization to supersymmetric theory, fermions are written in cohomological form. Different components of fermion transform as scalars and vector of $SO(4)$. Therefore the above results will suffice in determining their contribution.
\subsection*{Net multiplicity $N$}
It is easy to see that for round $S^4$ the net multiplicity can be found as
\bea
N_{S^4}&=&[S_e(m)+S_e(m+2)+S_e(m-2)]+[S_o(m)+S_o(m+2)+S_o(m-2)]-[V_e^0(m)+V_e^+(m)+V_e^-(m)]\nonumber\\
&-&[V_o^0(m)+V_o^+(m)+V_o^-(m)]\nonumber\\
&=& 2m
\eea
In the next step keeping in mind the  Dirichlet and Neumann BCs on the hemisphere given in section \ref{vectorBC}
\bea
N_{HS^4}^{Dir}&=&[2S_e(m)+S_o(m+2)+S_o(m-2)]-[V_e^0(m)+V_e^+(m)+V_e^-(m)]-S_o(m)\nonumber\\
&=&m+1
\eea
for Dirichlet BCs and 
\bea
N_{HS^4}^{Neum.}&=&[2S_0(m)+S_e(m+2)+S_e(m-2)]-[V_o^0(m)+V_o^+(m)+V_o^-(m)]-S_e(m)\nonumber\\
&=&m-1
\eea
for Neumann BCs.
Next since we know the eigenvalues of $\hat{Q}^2$, it is trivial to  write down the expressions for one-loop determinant.  However it is important to note that  there is some arbitrariness in the regularization scheme used here. For instance if one multiplies a factor of $e^{t C}$ for  constant $C\in \mathbb{R}$ , the net multiplicity $N_{HS^4}^{Dir}$ and 
$N_{HS^4}^{Neum.}$ is modified to $m+p$ and $m-p$ respectively, for some positive integer $p$, in such a way that $N_{HS^4}^{Neum.}+N_{HS^4}^{Dir}=2 m$.
\newpage
\section{Conclusions}\label{conclusion}
Despite extensive activity in Supersymmetric Localization computations on  curved manifolds in various dimensions, there were no first principle computations available  on hemisphere in four dimensions, although some educated guesses were given in \cite{Bullimore:2014nla}. We have done detailed computation  of wave functions on hemisphere $HS^4$ with supersymmetric BCs of Dirichlet  type, and also discussed briefly the Neumann BCs.  In the first part of this work, various one-loop determinants are computed using $SO(4)$ harmonics as the complete set of basis functions. The results obtained in the first part are re-checked in the second part where we do the same analysis in the framework of full $SO(5)$ harmonics. 
 We have also briefly discussed how the $N=2$ SUSY partition function on round $S^4$ $\grave{a}$ la Pestun \cite{Pestun:2007rz}, can be seen as composed of two Dirichlet type wave functions on southern and northern hemispheres properly glued together. The last observation can also be interpreted as kind of factorization of $Z_{S^4}$ wave function in terms of two hemisphere wave functions. Though this factorization should be seen as a convolution of two wave functions with non-trivial kernel, the later being the one-loop determinant of $N=2,3d$ gauge multiplet.
 \section{Acknowledgements}
 We would like to thank  Maszumi Honda for pointing out an important typo and Alejandro Cabo Bizet for insightful comments. NM would like to thank Bruno Le Floch for pointing out important issues in Instanton contributions discussed in section \ref{sec:joineds4} and for suggesting many improvements in the draft.
\newpage
\begin{appendices}
\section*{APPENDICES}
\section{Notation}
We use the same notation as in \cite{Hama:2012bg}. 
In the flat basis of the tangent space on $S^4$ we use the following set of Dirac  matrices ${\gamma^1,\gamma^2,\gamma^3,\gamma^4}$
\bea
\gamma^1&=&\left(
\begin{array}{cccc}
 0 & 0 & 0 & -i \\
 0 & 0 & -i & 0 \\
 0 & i & 0 & 0 \\
 i & 0 & 0 & 0 \\
\end{array}
\right), \gamma_2=\left(
\begin{array}{cccc}
 0 & 0 & 0 & -1 \\
 0 & 0 & 1 & 0 \\
 0 & 1 & 0 & 0 \\
 -1 & 0 & 0 & 0 \\
\end{array}
\right), \gamma_3=\left(
\begin{array}{cccc}
 0 & 0 & -i & 0 \\
 0 & 0 & 0 & i \\
 i & 0 & 0 & 0 \\
 0 & -i & 0 & 0 \\
\end{array}
\right), \nonumber\\\gamma_4&=&\left(
\begin{array}{cccc}
 0 & 0 & 1 & 0 \\
 0 & 0 & 0 & 1 \\
 1 & 0 & 0 & 0 \\
 0 & 1 & 0 & 0 \\
\end{array}
\right).
\eea
Pauli matrices are defined as usual
\bea
\tau^1= \left(
\begin{array}{cc}
 0 & 1 \\
 1 & 0 \\
\end{array}
\right),\tau^2= \left(
\begin{array}{cc}
 0 & -i \\
 i & 0 \\
\end{array}
\right),\tau^3= \left(
\begin{array}{cc}
 1 & 0 \\
 0 & -1 \\
\end{array}
\right)
\eea
In Weyl basis, with the decomposition $SO(4)\approx SU(2)_R\times SU(2)_L$ in chiral and anti-chiral basis, the sigma matrices $(\sigma^a)_{\alpha\dot{\beta}},(\bar{\sigma}^a)^{\dot{\alpha}\beta}$ are related to Pauli matrices as follows
\bea
\sigma^1=-i \left(
\begin{array}{cc}
 0 & 1 \\
 1 & 0 \\
\end{array}
\right),\sigma^2=-i \left(
\begin{array}{cc}
 0 & -i \\
 i & 0 \\
\end{array}
\right),\sigma^3=-i \left(
\begin{array}{cc}
 1 & 0 \\
 0 & -1 \\
\end{array}
\right),\sigma^4=-i \left(
\begin{array}{cc}
 1 & 0 \\
 0 & 1 \\
\end{array}
\right).
\eea
\bea
\bar{\sigma}^1=i \left(
\begin{array}{cc}
 0 & 1 \\
 1 & 0 \\
\end{array}
\right),\bar{\sigma}^2=i \left(
\begin{array}{cc}
 0 & -i \\
 i & 0 \\
\end{array}
\right),\bar{\sigma}^3=i \left(
\begin{array}{cc}
 1 & 0 \\
 0 & -1 \\
\end{array}
\right),\bar{\sigma}^4=i \left(
\begin{array}{cc}
 1 & 0 \\
 0 & 1 \\
\end{array}
\right).
\eea
The R-symmetry indices $A,B..$ and chiral and anti-chiral indices $\alpha,\dot{\alpha}$ are raised and lowered with antisymmetric matrices
$\epsilon^{\alpha\beta},\epsilon_{\alpha\beta},\epsilon^{\dot{\alpha}\dot{\beta}},\epsilon_{\dot{\alpha}\dot{\beta}},\epsilon_{AB},\epsilon^{AB}$ with the following matrix elements
\bea
\epsilon^{12}=1,\epsilon_{12}=-1,\epsilon^{\dot{1}\dot{2}}=1,\epsilon_{\dot{1}\dot{2}}=-1.
\eea

\section{$SO(4)\approx SU(2)_R\times SU(2)_L$ harmonics on $S^3$}\label{appendix:harmonics}
\subsection*{Background geometry}
The isometry group of round $S^4$ is $SO(5)$ and the most general way to compute the one-loop determinant is to use
$SO(5)$ harmonics. However since  we are interested in applying localization on a hemisphere, the boundary at $r=\frac{\pi}{2}$ breaks translational symmetry in $r$ coordinate, only $SO(4)\subset SO(5)$ is left intact and the best we can do is to use $SO(4)$ spherical harmonics.
We take the following metric on $S^4$
\bea
ds^2=g_{\mu\nu}dx^{\mu}dx^{\nu}=dr^2+\frac{sin(r)^2}{4}\big(d\theta^2+\sin\theta^2d\phi^2+(d\psi+\cos\theta d\phi)^2\big)
\eea
with coordinates $\mu=(\psi,\theta,\phi,r)$, 
such that   the  coordinates for Hopf fibration of unit $S^3$ part are:
\bea
z_1= \sin\bigg(\frac{\theta}{2}\bigg)e^{i\frac{(\psi-\phi)}{2}},z_2=\cos\bigg(\frac{\theta}{2}\bigg)e^{i\frac{(\psi+\phi)}{2}},
\eea
for $0\le\theta\le\pi$, $0\le\phi\le2\pi$ and $0\le\psi\le4\pi$.
As for the radial coordinate $r$,   $0\leq r \leq \pi$. The metric above describes $S^4$ as an $S^3$ fibration
over the $r$ interval$(0,\pi)$. $S^3$ has radius $sin(r)$ which vanishes at $0$ and $\pi$.
The vielbeins for $SU(2)_L$-frame and $SU(2)_R$-frame are
\bea
e^{\phantom{a}a}_{L\phantom{a}\mu}&=&
\left(
\begin{array}{cccc}
 0 & -\frac{1}{2} \cos (\psi ) sin(r) & -\frac{1}{2} sin(r) \sin (\theta ) \sin (\psi ) & 0 \\
 0 & \frac{1}{2} sin(r) \sin (\psi ) & -\frac{1}{2} \cos (\psi ) sin(r) \sin (\theta ) & 0 \\
 -\frac{sin(r)}{2} & 0 & -\frac{1}{2} \cos (\theta ) sin(r) & 0 \\
 0 & 0 & 0 & 1 \\
\end{array}
\right)
\eea
and
\bea
e^{\phantom{a}a}_{R\phantom{a}\mu}&=&\left(
\begin{array}{cccc}
 -\frac{1}{2} sin(r) \sin (\theta ) \sin (\phi ) & -\frac{1}{2} \cos (\phi ) sin(r) & 0 & 0 \\
 \frac{1}{2} \cos (\phi ) sin(r) \sin (\theta ) & -\frac{1}{2} sin(r) \sin (\phi ) & 0 & 0 \\
 -\frac{1}{2} \cos (\theta ) sin(r) & 0 & -\frac{sin(r)}{2} & 0 \\
 0 & 0 & 0 & 1 \\
\end{array}
\right)
\eea
 \subsection*{$SU(2)_L\times SU(2)_R$ Lie algebra}
The $SO(4)= SU(2)_R\times SU(2)_L$ Killing vectors  of $SU(2)_L$,
$J_L^a$ and  $SU(2)_R$,  $J_R^a$, $a=1,2,3$    are given
by $J_L^a=l^{a\mu}\partial_\mu$  and $J_R^a=r^{a\mu}\partial_\mu$, where:
\bea
l^{a\hspace{1mm}\mu}\equiv sin(r) \frac{i}{2}(1-2 \delta^{a 2})g^{\mu\nu}e^{\phantom{a}a}_{L\phantom{a}\nu},\qquad
r^{a\hspace{1mm}\mu}\equiv sin(r) \frac{i}{2}g^{\mu\nu}e^{\phantom{a}a}_{R\phantom{a}\nu}.
\eea

They obey the algebra:
\bea
[J_L^a,J_L^b]=i\epsilon_{abc}J_c^L,\quad [J_R^a,J_R^b]=i\epsilon_{abc}J_c^R,\quad [J_L^a,J_R^b]=0.
\eea

Notice that 
\bea
l^{3\hspace{1mm}\mu}=(-i ,0,0,0),\quad r^{3\hspace{1mm}\mu}=(0,0,-i ,0)
\eea
and
\bea
J^3_R e^{i(q_L\psi+q_R\phi)}&=&l^{3\hspace{1mm}\mu}\partial_{\mu}e^{i(q_L\psi+q_R\phi)}=
q_Le^{i(q_L\psi+q_R\phi)}
\nonumber\\
J^3_L e^{i(q_L\psi+q_R\phi)}&=&r^{3\hspace{1mm}\mu}\partial_{\mu}e^{i(q_L\psi+q_R\phi)}=
q_Re^{i(q_L\psi+q_R\phi)}
\eea
showing that $e^{i(q_L\psi+q_R\phi)}$ is an eigenfunction of $J^3_R$ and $J^3_L$. 
\subsection*{Scalar harmonics}
Highest weight states with respect to $SU(2)_L$ and $SU(2)_R$ for the scalar functions $ e^{i(j_L\psi+j_R\phi)} f(\theta)$    
are constructed by forming raising (lowering)
operators $J^\pm_L=J^1_L\pm i J^2_L$ and  $J^\pm_R=J^1_R\pm i J^2_R$. Highest weight states  are annihilated by $J^+_{L,R}$:
one can easily prove that this implies that $j_L=j_R$  and  $f(\theta)=\bigg(\cos\big(\frac{\theta}{2}\big)\bigg)^{2 j_L}$, up to a constant:
\bea
\Phi= e^{i(q_L\psi+q_R\phi)}\cos\bigg(\frac{\theta}{2}\bigg)^{2 j_L}
\eea
Applying lowering operators on $\Phi$ we can get all the other harmonics. In particular by applying $J^-_R$ $s$ times
we get states 
\bea
\Phi_s=e^{i(j_L\psi+(j_L-s)\phi)}\cos(\frac{\theta}{2})^{(2 jL-s)} \sin\bigg(\frac{\theta}{2}\bigg)^{s}
\eea
and one can check that when $s=2 j_L$ this is annihilated by $J^-_R$, i.e. it is a lowest weight state.\\
\subsection*{Vector harmonics}
Let us move on  to the vector harmonics: now we have to  consider Lie derivatives along the Killing vectors of $SO(4)$ 
acting on one-forms. The Lie derivative with respect to a vector field $K$ on a $1$-form $\omega$ is defined as
\bea
\mathcal{L}_X\omega\equiv (X,d\omega)+d(X,\omega)
\eea
In component form for $SU(2)_L\subset SO(4)$ 
\bea
\mathcal{L}^a_L\omega_{\nu}=l^{a\mu}(D_{\mu}\omega_{\nu}-D_{\nu}\omega_{\mu})+D_{\nu}(l^{a\mu}\omega_{\mu})
\eea
and for $SU(2)_R\subset SO(4)$ subgroup
\bea
\mathcal{L}^a_R\omega_{\nu}=r^{a\mu}(D_{\mu}\omega_{\nu}-D_{\nu}\omega_{\mu})+D_{\nu}(r^{a\mu}\omega_{\mu})
\eea
One can verify that Lie derivatives satisfy the Lie algebra relations: $[\mathcal{L}^a,\mathcal{L}^b]=\mathcal{L}^{[a,b]}$.
A basis of eigenstates of of the Cartan generators of both $SU(2)_{L,R}$ is given by:
\bea
\omega_{\nu}(\psi,\theta,\phi)=e^{i (q_L\psi+\q_R\phi)}\omega_{\nu}^1(\theta)
\eea
Now for  $q_L=j_L,q_R=j_R$, this will be a highest weight state if it is annihilated by the raising operators of $SU(2)_R\times SU(2)_L$ Lie algebra
\bea
\mathcal{L}_L^1\omega_{\nu}+i \mathcal{L}_L^2\omega_{\nu}=0\nonumber\\
\mathcal{L}_R^1\omega_{\nu}+i \mathcal{L}_R^2\omega_{\nu}=0
\eea
Solving these differential equations gives the following general solution:
\bea
\omega_1^{\psi}(\theta)&=&\alpha_{1} \sin\left(\frac{\theta }{2}\right) ^{-\text{jL}} \cos \left(\frac{\theta }{2}\right)^{\text{j}_L} \sin (\theta )^{\text{j}_R},\quad \omega_1^{\phi}(\theta)=\alpha_{2} \sin (\theta )^{\text{j}_L} \sin \left(\frac{\theta }{2}\right)^{-\text{j}_R} \cos \left(\frac{\theta }{2}\right)^{\text{j}_R}, \nonumber\\
\omega_1^{\theta}(\theta)&=&\frac{1}{2} i \sec \left(\frac{\theta }{2}\right) \sin \left(\frac{\theta }{2}\right)^{-\text{j}_L-\text{j}_R-1} \left(\alpha_{2} \sin \left(\frac{\theta
   }{2}\right)^{\text{j}_L} \sin(\theta ) ^{\text{j}_L} \cos \left(\frac{\theta }{2}\right)^{\text{j}_L}\right)
\nonumber\\
&-&\left(\alpha_{1} \cos \left(\frac{\theta }{2}\right)^{\text{j}_L} \sin
  \left(\frac{\theta }{2}\right) ^{\text{j}_R} \sin (\theta )^{\text{j}_R} (\cos (\theta ) (\text{j}_L-\text{j}_R+1)+\text{j}_L-\text{j}_R)\right).
\eea
along with the following three choices of  constant parameters\\
a)$j_R=j_L,\alpha_{1}\neq 0,\alpha_{2}= \alpha_{1}$\\
b)$j_R=j_L+1,\alpha_{1}\neq 0,\alpha_{2}= 0$.\\
c)$j_R=j_L-1,\alpha_{1}= 0$.\\
In all these cases the solutions are regular at the North and South Poles, $r=0$ and $r=\pi$ respectively. \\
These results  agree of course  with the fact that the $(j_L,j_R)$ representation
of $SU(2)_R\times SU(2)_L$  contains the $j=1$ representation of the diagonal $SU(2)$, the one acting on the tangent space of $S^3$, if
and only if $|j_L-j_R| \leq 1$ and the scalar $j=0$ if and only if $|j_L-j_R| =0$.\\ 
\subsection*{Projection to $SU(2)_L$ basis}
The analysis simplifies if we  work with the $SO(3)$ tangent space basis, i.e. convert world indices of gauge fields to tangent ones by contracting with $l^{a\mu}$:
\bea
\omega_{f\hspace{1mm}}^a=l^{a\mu}\omega_{\mu}
\eea
In this basis the gauge fields behave like scalars and will always be in the $j_L=j_R$ representation of $SU(2)_L\times SU(2)_R$ , as can be explicitly verified.



Now we specialize  gauge field $A_\mu$ , $\mu=1,2,3$, to the flat basis $A_a=l^{a\mu}A_{\mu}$ where $ a=1,2,3$ or $a=+,-,3$ where $A_+\equiv A_1+i A_2, A_-\equiv A_1-i A_2$ and similarly $\chi_+\equiv\chi_1+i \chi_2, \chi_-\equiv\chi_1-i \chi_2$
Taking into account the relation between $\hat{Q}^2$ eigenvalues and the $SU(2)_L$ wights $q_L$, including the shifts in the weights, we are led to the following Fourier expansions.\\
Fields in the kernel of $D^{vec}_{10}$:

\bea
A_+(\psi,\theta,\phi,r)&=&e^{-i(q_L-1)\psi}e^{-i q_R\phi}A_+(\theta,r),\quad
A_-(\psi,\theta,\phi,r)=e^{-i(q_L+1)\psi}e^{-i q_R\phi}A_-(\theta,r)\nonumber\\
A_r(\psi,\theta,\phi,r)&=&e^{-i(q_L)\psi}e^{-i q_R\phi}A_4(\theta,r),\quad
A_3(\psi,\theta,\phi,r)=e^{-i(q_L)\psi}e^{-i q_R\phi}A_3(\theta,r)\nonumber\\
\phi_2(\psi,\theta,\phi,r)&=&e^{-i(q_L)\psi}e^{-i q_R\phi}\phi_1(\theta,r).\nonumber\\
\eea
Fields in the cokernel of $D^{vec}_{10}$:
\bea\label{eq:FourierCok1}
\chi_+(\psi,\theta,\phi,r)&=&e^{i(q_L+1)\psi}e^{i q_R\phi}\chi_+(\theta,r),\quad
\chi_-(\psi,\theta,\phi,r)=e^{i(q_L-1)\psi}e^{i q_R\phi}\chi_-(\theta,r)\nonumber\\
\chi_3(\psi,\theta,\phi,r)&=&e^{i(q_L)\psi}e^{i q_R\phi}\chi_3(\theta,r),\quad
\text{c}(\psi,\theta,\phi,r)=e^{i(q_L)\psi}e^{i q_R\phi}\text{c}(\theta,r)\nonumber\\
\tilde{c}(\psi,\theta,\phi,r)&=&e^{i(q_L)\psi}e^{i q_R\phi}\tilde{\text{c}}(\theta,r).\nonumber\\
\eea
To write the kernel equations in a more suggestive form, we  redefine the $\phi_1(\theta,r)$ field as
\bea
\phi_1(\theta,r)\equiv \frac{1}{2}(-2 i A_3(\theta,r-\Lambda(\theta,r))\sec(r).
\eea
where $\Lambda(\theta,r))$ is field dependent gauge transformation.
\subsection*{Regularity at North and South poles}
At the two poles of $S^4$ the space locally looks like  $R^4$ and to check the regularity of $A_{\mu}$ one has to expand its components as polynomials  in $z_1,z_2,\bar{z}_1,\bar{z}_2$ and find the leading behavior in the limit of  $z_1\to 0,z_2\to 0,\bar{z}_1\to 0,\bar{z}_2\to 0$ or in terms of polar variable $r\to0$. It is easy to find that the highest weight state with $q_L=j_L$ and $q_R=j_R$ and because of expansion in scalar harmonics $j_L=j_R$
\bea
A_r&\sim& r^{2j_L-1}, \quad for \quad j_L=j_R\ne 0,\nonumber\\
A_r&\sim& r \quad for \quad j_L=j_R= 0
\eea
whereas for $A_i$ with $i=\psi,\theta,\psi$ one gets the following leading order behavior
\bea
A_i&\sim& r^{2j_L}, \quad for \quad j_L=j_R\ne 0,\nonumber\\
A_i&\sim& r^2 \quad for \quad j_L=j_R= 0
\eea
\\ 
However in tangent space basis all the components of gauge field $A_a=e_a^{\mu}A_{\mu}$ with $a=1,2,3,4$ have identical leading behavior
\bea
A_a&\sim& r^{2j_L-1}, \quad for \quad j_L=j_R\ne 0,\nonumber\\
A_a&\sim& r \quad for \quad j_L=j_R= 0.
\eea
The consequences of these regularity properties are analyzed in detail in section \ref{vectoranalysis} and in appendix \ref{appendix:ode4}. 
Since for computational purposes fermions are written in terms of cohomological variables, their regularity behavior can be easily deduced from that of the gauge field $A_\mu$ given above.
\section{$N=2$ off-shell  SUSY transformations}\label{appendix:susytr}
For completeness we reproduce here the supersymmetric transformation rules of vector and matter multiplet for general background auxiliary fields.
\subsection{vector multiplet }
\bea
QA_\m&=&i\xi^A\sigma_\m\bar{\lambda}_A-i\bar{\xi}^A\bar{\sigma}_\m\lambda_A,\nonumber\\
Q\phi&=&-i\xi^A\lambda_A,\qquad
Q\bar{\phi}=i \bar{\xi}^A\bar{\lambda}_A,\nonumber\\
Q\lambda_A&=& \frac{1}{2}\sigma^{\m\n}\xi_A(F_{\m\n}+8\bar{\phi}T_{\m\n})+2\sigma^\m\bar{\xi}_A D_\m\phi+
\sigma^\m D_\m\bar{\xi}_A\phi+2 i \xi_A[\phi,\bar{\phi}]+D_{AB}\xi^B,\nonumber\\
Q\bar{\lambda}_A&=&\frac{1}{2}\bar{\sigma}^{\m\n}\bar{\xi}_A(F_{\m\n}+8\phi\bar{T}_{\m\n})+2\bar{\sigma}^\m\xi_A D_\m\bar{\phi}+\bar{\sigma}^\m D_\m\xi_A\bar{\phi}-2 i \bar{\xi}_A[\phi,\bar{\phi}]+D_{AB}\bar{\xi}^B,\nonumber\\
Q D_{AB}&=&-i\bar{\xi}_A\bar{\sigma}^\m D_\m\lambda_B-i\bar{\xi}_B\bar{\sigma}^\m D_\m\lambda_A+i\xi_A\sigma^\m D_\m\bar{\lambda}_B+i\xi_B\sigma^\m D_\m\bar{\lambda}_A\nonumber\\
&-&2[\phi,\bar{\xi}_A\bar{\lambda}_B+\bar{\xi}_B\bar{\lambda}_A]+2[\bar{\phi},\xi_A\lambda_B+\xi_B\lambda_A].
\eea
\subsection{matter multiplet }
\bea
Q q_A&=&-i\xi_A\psi +i\bar{\xi}_A\bar{\psi},\nonumber\\
Q \psi&=& 2\sigma^\m\bar{\xi}_A D_\m q^A+\sigma^\m D_\m\bar{\xi}_Aq^A-4i\xi_A\bar{\phi}q^A+2\check{\xi}_AF^A,\nonumber\\
Q \bar{\psi}&=& 2\bar{\sigma}^\m\xi_A D_\m q^A+\bar{\sigma}^\m D_\m \xi_Aq^A-4i\bar{\xi}_A\phi q^A+2\bar{\check{\xi}}_AF^A,\nonumber\\
Q F_A&=&i\check{\xi}_A\sigma^\m D_\m\bar{\psi}-2\check{\xi}_A\phi\psi-2\check{\xi}_A\lambda_B q^B+2i\check{\xi}_A(\sigma^{\m\n}T_{\m\m})\psi\nonumber\\
&-&i\bar{\check{\xi}}_A\bar{\sigma}^{\m}D_\m\psi+2\bar{\check{\xi}}_A\bar{\phi}\bar{\psi}+2\bar{\check{\xi}}_A\bar{\lambda}_Bq^B-2i\bar{\check{\xi}}_A(\bar{\sigma}^{\m\n}\bar{T}_{\m\n})\bar{\psi}.
\eea
\section{Differential equations}\label{KernelE}
\subsection*{Kernel}
Varying $V_{vec}$ with respect to $\chi_3,\chi_+,\chi_-,\text{c}$ and $\bar{c}$ respectively generates  following differential equations for the fields belonging to kernel of $D_{10}$:
\bea
\mathcal{E}_1&=&\frac{1}{2} \left(4 \sin (\theta ) \tan (r) \partial_{\theta}\text{A}_3(\theta ,r)-4 \text{q}_L \tan (r) \text{A}_3(\theta ,r)+4 \text{q}_R \cos (\theta ) \tan (r)
   \text{A}_3(\theta ,r)\right)\nonumber\\&-&\left(\sin (\theta ) \sin ^2(r) \partial_{\theta}\text{A}_r(\theta ,r)+\sin ^2(r) \text{A}_r(\theta ,r) (\text{q}_L-\text{q}_R \cos (\theta ))+2
   i \sin (\theta ) \sin ^2(r) \partial_r\text{A}_+(\theta ,r)\right)\nonumber\\&+&\left(2 i \text{q}_R \sin (\theta ) \sin (2 r) \text{A}_+(\theta ,r)-2 i \sin (\theta ) \sin ^2(r)
   \tan (r) \partial_{\theta}\Lambda(\theta ,r)+2 i \text{q}_L \sin ^2(r) \tan (r) \Lambda(\theta ,r)\right)\nonumber\\&-&\left(2 i \text{q}_R \cos (\theta ) \sin ^2(r) \tan (r)
   \Lambda(\theta ,r)\right),
\nonumber\\
\mathcal{E}_2&=&\frac{1}{2} \tan (r) \left(i \left(2 \sin ^2(r) \Lambda(\theta ,r) (\text{q}_L-\text{q}_R \cos (\theta ))+\sin (\theta ) \left(4 i
   \partial_{\theta}\text{A}_3(\theta ,r)\right)+\left(i \sin (r) \cos (r) \partial_{\theta}\text{A}_r(\theta ,r)\right)\right)\right)\nonumber\\&+&\left(\left(\left(\sin (2 r) \partial_r\text{A}_-(\theta ,r)-4 \text{q}_R \cos ^2(r)
   \text{A}_-(\theta ,r)\right)+\left(2 \sin ^2(r) \partial_{\theta}\Lambda(\theta ,r)\right)\right)-4\left( \text{A}_3(\theta ,r) (\text{q}_L-\text{q}_R \cos (\theta ))\right)\right)\nonumber\\&+&\left(\sin
   (r) \cos (r) \text{A}_r(\theta ,r) (\text{q}_R \cos (\theta )-\text{q}_L)\right),
\nonumber\\
\mathcal{E}_3&=&\frac{1}{4} \sin (r) \cos (r) \left(\sin (\theta ) \left(\sec (r) \left(8 i \tan (r) \partial_r\text{A}_3(\theta ,r)-8 \partial_{\theta}\text{A}_-(\theta ,r)+8
   \partial_{\theta}\text{A}_+(\theta ,r)\right)\right)\right)\nonumber\\&+&\left(\left(\left(3 \tan (r) \partial_r\Lambda(\theta ,r)-\sin (3 r) \sec (r) \partial_r\Lambda(\theta ,r)+4 \tan ^2(r)
   \Lambda(\theta ,r)\right)+4 i (\cos (2 r)+3) \sec ^3(r) \text{A}_3(\theta ,r)\right)\right)\nonumber\\&+&\left(\left(4 i \text{q}_R \sin (r) \text{A}_r(\theta ,r)\right)-8 \sec (r)
   \text{A}_-(\theta ,r) ((\text{q}_R+1) \cos (\theta )-\text{q}_L)+8 \sec (r) \text{A}_+(\theta ,r) \cos (\theta )\right)\nonumber\\&+&\left(\text{q}_L-\text{q}_R \cos (\theta)\right),
\nonumber\\
\mathcal{E}_4&=&-\frac{1}{4} \sin (r) \left(\sin (\theta ) \left(16 \text{q}_R \text{A}_3(\theta ,r)+\partial_{r}\text{A}_r(\theta ,r)-\cos (2 r) \partial_r\text{A}_r(\theta
   ,r)+3 \sin (2 r) \text{A}_r(\theta ,r)\right)\right)\nonumber\\&+&\left(8 i \partial_{\theta}\text{A}_-(\theta ,r)+8 i \partial_{\theta}\text{A}_+(\theta ,r)\right)+8 i \text{A}_-(\theta ,r)
   ((\text{q}_R+1) \cos (\theta )-\text{q}_L)+8 i \text{A}_+(\theta ,r) (\cos (\theta )\nonumber\\&+&\text{q}_L-\text{q}_R \cos (\theta )),\nonumber\\
\mathcal{E}_5&=&\frac{1}{2} \left(4 \sin (\theta ) \tan (r) \partial_{\theta}\text{A}_3(\theta ,r)-4 \text{q}_L \tan (r) \text{A}_3(\theta ,r)
+4 \text{q}_R \cos (\theta ) \tan (r)
   \text{A}_3(\theta ,r)\right)\nonumber\\
   &-&\left(\sin (\theta ) \sin ^2(r) \partial_{\theta}\text{A}_r(\theta ,r)
   +\sin ^2(r) \text{A}_r(\theta ,r) (q_L-q_R \cos (\theta ))+2
   i \sin (\theta ) \sin ^2(r) \partial_r\text{A}_+(\theta ,r)\right)\nonumber\\
   &+&\left(2 i \text{q}_R \sin (\theta ) \sin (2 r) \text{A}_+(\theta ,r)-2 i \sin (\theta ) \sin ^2(r)
   \tan (r) \partial_{\theta}\Lambda(\theta ,r)+2 i \text{q}_L \sin ^2(r) \tan (r) \Lambda(\theta ,r)\right)\nonumber\\
   &-&\left(2 i \text{q}_R \cos (\theta ) \sin ^2(r) \tan (r)
   \Lambda(\theta ,r)\right).
\eea
\subsection*{Cokernel}
Similar to  kernel equations, to get the zero mode differential equations for co-kernel fields one has to vary $V_{vec}$ with respect to $A_+,A_-,A_3,A_4$ and $\Lambda$ respectively to generate the following:
\bea
\mathcal{CE}_1&=&\frac{1}{8} \sin (r) \left(\sin (\theta ) \left(4 \text{q}_L \partial_{\theta}\text{c}(\theta ,r)+2 i \partial_{\theta}\bar{c}(\theta ,r)-2 i (\text{q}_L-1) \cos (r)
   \chi_-(\theta ,r)\right)\right)\nonumber\\&+&\left(\left(2 \partial_{\theta}\chi_3(\theta ,r)+i \sin (r) \partial_r\chi_-(\theta ,r)\right)-2 \text{q}_R (2 \text{q}_L
   \text{c}(\theta ,r)+i \bar{c}(\theta ,r)+\chi_3(\theta ,r))\right)\nonumber\\&+&\left(2 \text{q}_L \cos (\theta ) (2 \text{q}_L \text{c}(\theta ,r)+i
   \bar{c}(\theta ,r)+\chi_3(\theta ,r))\right),\nonumber\\
 \mathcal{CE}_2&=&\frac{1}{8} \sin (r) \left(i \sin (\theta ) \left(-4 i \text{q}_L \partial_{\theta}\text{c}(\theta ,r)+2 \partial_{\theta}\bar{c}(\theta ,r)+2 (\text{q}_L+1) \cos (r)
   \chi_+(\theta ,r)\right)\right)\nonumber\\&+&\left(\left(2 i \partial_{\theta}\chi_3(\theta ,r)+\sin (r) \partial_r\chi_+(\theta ,r)\right)+4 \text{q}_L \text{c}(\theta
   ,r) (\text{q}_R-\text{q}_L \cos (\theta ))\right)\nonumber\\&+&\left(2 i \bar{c}(\theta ,r) (\text{q}_R\-\text{q}_L \cos (\theta ))+2 \text{q}_L \cos (\theta ) 
   \chi_3(\theta ,r)-2 \text{q}_R \chi_3(\theta ,r)\right),\nonumber\\
\mathcal{CE}_3&=&\frac{1}{4} \tan (r) \left(\sin (\theta ) \left(2 \text{q}_L \cos (r) (-\bar{c}(\theta ,r)+2 i \text{q}_L \text{c}(\theta ,r))+i \sin (r) \partial_r
\chi_3(\theta ,r)\right)\right)\nonumber\\&+&\left(\left(\partial_{\theta}\chi_-(\theta ,r)-\partial_{\theta}\chi_+(\theta ,r)\right)-\cos (\theta ) (\text{q}_L \chi_-(\theta ,r)+
(\text{q}_L+1) \chi_+(\theta ,r))\right)\nonumber\\&+&\left((\cos (\theta )+\text{q}_R) \chi_-(\theta ,r)+\text{q}_R \chi_+(\theta
   ,r)\right),\nonumber\\
\mathcal{CE}_4&=&\frac{1}{8} \sin ^2(r) \left(\sin (\theta ) \left(-2 i \text{q}_L \left(\sin (r) \partial_r\text{c}(\theta ,r)+\cos (r) \chi_3(\theta
   ,r)\right)+\sin (r) \partial_r\bar{c}(\theta ,r)\right)\right)\nonumber\\&-&\left(\left(\partial_{\theta}\chi_-(\theta ,r)-\partial_{\theta}\chi_+(\theta ,r)\right)+\cos (\theta )
   ((\text{q}_L-1) \chi_-(\theta ,r)-(\text{q}_L+1) \chi_+(\theta ,r))\right)\nonumber\\&+&\left(\text{q}_R (\chi_+(\theta ,r)-\chi_-(\theta
   ,r))\right),\nonumber\\
\mathcal{CE}_5&=&-\frac{1}{16} \sin (r) \left(\sin (\theta ) \left(2 \sin ^2(r) \left(\partial_r^2\text{c}(\theta ,r)+2 i \text{q}_L \bar{c}(\theta ,r)-3 
\chi_3(\theta ,r)\right)+8 \partial_{\theta}^2\text{c}(\theta ,r)\right)\right)\nonumber\\&+&\left(\left(3 \sin (2 r) \partial_r\text{c}(\theta ,r)\right)+8 \cos (\theta ) \partial_{\theta}\text{c}(\theta
   ,r)-8 \text{c}(\theta ,r) \left(\csc (\theta ) \left(\text{q}_L^2\right)\right)\right)\nonumber\\&+&\left(\left(\left(\text{q}_R^2\right)-\text{q}_L^2 \sin (\theta ) \sin ^2(r)-2 \text{q}_L \text{q}_R
   \cot (\theta )\right)\right) .     
\eea
\section{Localizing fermionic functional $V_{hyper}$}\label{appendix:fermfunctional}
Instead of writing down the kernel and co-kernel differential equations for the hyper multiplet, we only provide the fermionic functional  in cohomological variables is given 
\bea
V_{hyper}&=&\frac{1}{64} \sin ^2(r) \bigg[\left(2 \left(2 \sin (\theta ) \partial_{\theta}\text{q}_{11}(\theta ,r)
   \Sigma_{12}(\theta ,r)-i \sin (\theta ) \sin (r) \partial_r\text{q}_{11}(\theta ,r)
   \Sigma_{22}(\theta ,r)\right)\right)\nonumber\\&+&\left(\left(\text{q}_{11}(\theta ,r) (\Sigma_{12}(\theta ,r) (\cos (\theta )+2
   \text{q}_L \cos (\theta )-2 \text{q}_R)-i (2 \text{q}_L+1) \sin (\theta ) \cos (r)
   \Sigma_{22}(\theta ,r))\right)\right)\nonumber\\&-&\left(\left(2 \sin (\theta ) \partial_{\theta}\text{q}_{12}(\theta ,r)
   \Sigma_{11}(\theta ,r)+i \sin (\theta ) \sin (r) \partial_r\text{q}_{12}(\theta ,r)
   \Sigma_{21}(\theta ,r)\right)\right)\nonumber\\&-&\left(\left(i \sin (\theta ) \sin (r) \partial_r\text{q}_{21}(\theta ,r)
   \Sigma_{12}(\theta ,r)+2 \sin (\theta ) \partial_{\theta}\text{q}_{21}(\theta ,r) \Sigma_{22}(\theta
   ,r)\right)\right)\nonumber\\&+&2\left(\left( i \text{q}_L \sin (\theta ) \cos (r) \text{q}_{21}(\theta ,r) \Sigma_{12}(\theta
   ,r)-2 \text{q}_L \cos (\theta ) \text{q}_{21}(\theta ,r) \Sigma_{22}(\theta ,r)\right)\right)\nonumber\\&+&2\left(\left( \text{q}_R
   \text{q}_{21}(\theta ,r) \Sigma_{22}(\theta ,r)-i \sin (\theta ) \cos (r)
   \text{q}_{21}(\theta ,r) \Sigma_{12}(\theta ,r)+\cos (\theta ) \text{q}_{21}(\theta ,r)
   \Sigma_{22}(\theta ,r)\right)\right)\nonumber\\&+&\left(\left(i \sin (\theta ) \sin (r) \partial_r\text{q}_{22}(\theta ,r)
   \Sigma_{11}(\theta ,r)-2 \sin (\theta ) \partial_{\theta}\text{q}_{22}(\theta ,r) \Sigma_{21}(\theta
   ,r)\right)\right)\nonumber\\&+&\left(\left(\text{q}_{22}(\theta ,r) (\Sigma_{21}(\theta ,r) (-(\cos (\theta )-2 \text{q}_L \cos
   (\theta )+2 \text{q}_R))-i (2 \text{q}_L-1) \sin (\theta ) \cos (r) \Sigma_{11}(\theta
   ,r))\right)\right)\nonumber\\&+&\left(\text{q}_{12}(\theta ,r) (-2 \Sigma_{11}(\theta ,r) (\cos (\theta )+2
   \text{q}_L \cos (\theta )-2 \text{q}_R)+2 i (2 \text{q}_L+1) \sin (\theta ) \cos (r)
   \Sigma_{21}(\theta ,r))\right)\bigg].\nonumber\\
\eea
\section{Analysis of kernel and cokernel equations for  $j_L=0$ }
\subsection*{Hemisphere}

With a boundary at $r=\frac{\pi}{2}$ the regular solutions of kernel equations, with the unique choice $q_L=0,q_R=0$, are 
\bea
\text{A}_r^{(0,0,0)}(r)&=&-\frac{1}{12} \text{b}_0 \left(3 \sin \left(\frac{r}{2}\right)+\sin \left(\frac{3 r}{2}\right)\right) \sec ^3\left(\frac{r}{2}\right),\nonumber\\
\text{A}_3^{(0,0,0)}(r)&=&\frac{1}{288} i \left(\csc ^2(r) \left(\cos (2 r) \left(-24 \bar{a}_0 \log \left(\sin \left(\frac{r}{2}\right)\right)+24 \bar{a}_0 \log (\sin
   (r))\right)\right)\right)\nonumber\\&+& \left(24 \bar{a}_0 \log \left(\cos \left(\frac{r}{2}\right)\right)-61 \bar{a}_0+72 \text{d}_0\right)+2 \bar{a}_0 \cos (4 r)-72 \bar{a}_0 \log
   \left(\sin \left(\frac{r}{2}\right)\right)\nonumber\\&+&72 \left(\bar{a}_0 \log (\sin (r))+72 \bar{a}_0 \log \left(\cos \left(\frac{r}{2}\right)\right)-96
   \bar{a}_0 \cos (r) \log \left(\sec ^2\left(\frac{r}{2}\right)\right)-57 \bar{a}_0+216 \text{d}_0\right)\nonumber\\&-&\frac{4 \sqrt{\cos (r)} \bar{a}_0 (24
   \log (2)-29)+72 d_0 \sqrt{\sin (r) \cos (r)}}{\sin ^{\frac{5}{2}}(r)},\nonumber\\
\Lambda^{(0,0,0)}(r)&=& \frac{1}{24} \left(-9 \bar{a}_0 \cot ^2(r)+2 \bar{a}_0 \csc ^2\left(\frac{r}{2}\right)+\bar{a}_0 \csc ^2(r)-2 \bar{a}_0 \sec
   ^2\left(\frac{r}{2}\right)\right)\nonumber\\&-&8\left( \bar{a}_0 \log \left(\sin \left(\frac{r}{2}\right)\right)+8 \bar{a}_0 \log (\sin (r))+8 \bar{a}_0 \log \left(\cos
   \left(\frac{r}{2}\right)\right)+24 d_0\right).
\eea 
Similarly cokernel equations can be solved easily for $j_L=0$ to give following regular solutions 
\bea
\chi_3^{(0,0,0)}(r)&=&\frac{C_{1}}{2},\quad \bar{c}^{(0,0,0)}(r)=C_2,\nonumber\\
c^{(0,0,0)}(r)&=&\frac{1}{16} \left(9 C_1 \cot ^2(r)-2 C_{1} \csc ^2\left(\frac{r}{2}\right)-C_{1} \csc ^2(r)+2 C_{1} \sec
   ^2\left(\frac{r}{2}\right)+8 C_{1} \log \left(\sin \left(\frac{r}{2}\right)\right)\right)\nonumber\\&-&8\left( C_{1} \log (\sin (r))
   -8 C_{1} \log \left(\cos
   \left(\frac{r}{2}\right)\right)+16 C_{0}\right).
\eea
where $b_0,\bar{a}_0,d_0,C_0,C_1,C_2$ are constant functions.\\
However with Dirichlet BCs $A_r|_{r=\frac{\pi}{2}}=0,A_3|_{r=\frac{\pi}{2}}=0$ and $\phi_2|_{r=\frac{\pi}{2}}=0$ imposed at the boundary for the fluctuation fields and by the requirement that  these BCs should be closed under supersymmetry, we get that all of the constants  $b_0,\bar{a}_0,d_0,C_0,C_1,C_2$ must vanish.

\section{ Identities used}\label{identities}
\subsection{For kernel}
For simplicity we take the basis for the harmonics as $e^{i(q_L\psi+q_R\phi)}Y^{(j_L,q_L,q_R)}(\theta)$ satisfying
\bea
 J^{-}(e^{i(q_L\psi+q_R\phi)}Y^{(j_L,q_L,q_R)}(\theta))=e^{i((q_L-1)\psi+q_R\phi)}Y^{(j_L,q_L-1,q_R)}(\theta)
\eea
and 
\bea
 J^{+}(e^{i(q_L\psi+q_R\phi)}Y^{(j_L,q_L,q_R)}(\theta))=(j_L-q_L)(j_L+q_L+1)e^{i((q_L+1)\psi+q_R\phi)} Y^{(j_L,q_L+1,q_R)}(\theta)
\eea
This basis is not normalized but it is irrelevant for the present analysis.
It is trivial to see the following identities hold
\bea
Y^{(j_L,j_L+1,-q_R)}(\theta)&=& 0,\qquad Y^{(j_L,-j_L-1,-q_R)}(\theta)= 0,\nonumber\\
A_+^{(j_L,j_L+1,-q_R)}(r)&=& 0,\qquad A_+^{(j_L,-j_L-1,-q_R)}(r)= 0,\nonumber\\
A_-^{(j_L,j_L+1,-q_R)}(r)&=& 0,\qquad A_-^{(j_L,-j_L-1,-q_R)}(r)= 0,\nonumber\\
A_+^{(j_L,j_L+2,-q_R)}(r)&=& 0,\qquad A_+^{(j_L,-j_L-2,-q_R)}(r)= 0,\nonumber\\
A_-^{(j_L,j_L+2,-q_R)}(r)&=& 0,\qquad A_-^{(j_L,-j_L-2,-q_R)}(r)= 0,\nonumber\\
A_3^{(j_L,j_L+1,-q_R)}(r)&=& 0,\qquad A_3^{(j_L,-j_L-1,-q_R)}(r)= 0,\nonumber\\
A_4^{(j_L,j_L+1,-q_R)}(r)&=& 0,\qquad A_4^{(j_L,-j_L-1,-q_R)}(r)= 0.
\eea
To evalute the Kernel equations for different $SU(2)_R$ charges we need the following identities.\\
For $\mathcal{E}_1,\mathcal{E}_2$
\bea
\partial_{\theta}Y^{(j_L,q_L+1,q_R)}(\theta)&=i Y^{(j_L,q_L,q_R)}(\theta)-(-q_R+(1+q_L)\cos(\theta))\csc(\theta)\nonumber\\ 
&\times Y^{(j_L,q_L+1,q_R)}(\theta),\nonumber\\
\partial_{\theta}Y^{(j_L,q_L-1,q_R)}(\theta)&=-(q_R+\cos(\theta)(1-q_L))\csc(\theta)Y^{(j_L,q_L-1,q_R)}(\theta)+\nonumber\\
& i (1+j_L-q_L)(j_L+q_L)s Y^{(j_L,q_L,q_R)}(\theta).
\eea
For $\mathcal{E}_3$
\bea
\partial_{\theta}Y^{(j_L,q_L,q_R)}(\theta)&=i Y^{(j_L,q_L-1,q_R)}(\theta)+(-q_L\cot(\theta)+q_R\csc(\theta))\nonumber\\
&\times Y^{(j_L,q_L,q_R)}(\theta).
\eea
and for $\mathcal{E}_4$
\bea
\partial_{\theta}Y^{(j_L,q_L,q_R)}(\theta)&=(-q_R+q_L\cos(\theta))\csc(\theta)Y^{(j_L,q_L,q_R)}(\theta)+
\nonumber\\
& i (j_L-q_L)(1+j_L+q_L)Y^{(j_L,1+q_L,q_R)}(\theta).
\eea

\subsection{For Cokernel}
We need the following relations
\bea
Y^{(j_L,j_L+1,q_R)}(\theta)&=&0,\quad Y^{(j_L,-j_L-1,q_R)}(\theta)=0,\nonumber\\
\chi_+^{(j_L,j_L+1,q_R)}(r)&=&0,\quad \chi_+^{(j_L,-j_L-1,q_R)}(r)=0,\nonumber\\
\chi_-^{(j_L,j_L+1,q_R)}(r)&=&0,\quad \chi_-^{(j_L,-j_L-1,q_R)}(r)=0,\nonumber\\
\chi_+^{(j_L,j_L+2,q_R)}(r)&=&0,\quad \chi_+^{(j_L,-j_L-2,q_R)}(r)=0,\nonumber\\
\chi_-^{(j_L,j_L+2,q_R)}(r)&=&0,\quad \chi_-^{(j_L,-j_L-2,q_R)}(r)=0,\nonumber\\
\chi_3^{(j_L,j_L+1,q_R)}(r)&=&0,\quad \chi_3^{(j_L,-j_L-1,q_R)}(r)=0,\nonumber\\
c_{a_{1}}^{(j_L,j_L+1,q_R)}(r)&=&0,\quad c_{a_{1}}^{(j_L,-j_L-1,q_R)}(r)=0,\nonumber\\
\text{c}^{(j_L,j_L+1,q_R)}(r)&=&0,\quad \text{c}^{(j_L,-j_L-1,q_R)}(r)=0.
\eea
Different useful identities involving derivative of harmonics are, for $C\mathcal{E}_1,C\mathcal{E}_2$
\bea
\partial_{\theta}Y^{(j_L,q_L+1,q_R)}(\theta)&=&i Y^{(j_L,q_L,q_R)}(\theta)-(-q_R+(1+q_L)\cos(\theta))\csc(\theta)\nonumber\\
&\times& Y^{(j_L,q_L+1,q_R)}(\theta),\nonumber\\
\partial_{\theta}Y^{(j_L,q_L-1,q_R)}(\theta)&=&i(j_L+q_L)(1+j_L-q_L) Y^{(j_L,q_L,q_R)}(\theta)\nonumber\\
&-&(q_R+\cos(\theta)(1-q_L))\csc(\theta) Y^{(j_L,q_L-1,q_R)}(\theta).
\eea
for $C\mathcal{E}_3$
\bea
\partial_{\theta}Y^{(j_L,q_L,q_R)}(\theta)&=&i Y^{(j_L,q_L-1,q_R)}(\theta)+(-q_L\cot(\theta)\nonumber\\
&+& q_R\csc(\theta))Y^{(j_L,q_L,q_R)}(\theta),
\eea
for $C\mathcal{E}_4$
\bea
\partial_{\theta}Y^{(j_L,q_L,q_R)}(\theta)&=&i (j_L-q_L)(1+j_L+q_L) Y^{(j_L,q_L+1,q_R)}(\theta)\nonumber\\
&+&(q_L\cos(\theta)-q_R)\csc(\theta)Y^{(j_L,q_L,q_R)}(\theta),
\eea
and for $C\mathcal{E}_5$
\bea
\partial^2_{\theta}Y^{(j_L,q_L,q_R)}&=&-(j_L(1+ j_L)+2 q_L q_R\cot(\theta)\csc(\theta)\nonumber\\
&-&(q_L^2+q_R^2)\csc(\theta)^2)Y^{(j_L,q_L,q_R)}-\cot(\theta)\partial_{\theta}Y^{(j_L,q_L,q_R)}.
\eea
\section{Analysis for $|q_L|<j_L$}\label{appendix:ode4}
For $q_L=0$
\bea
A_4^{(j_L,0,-q_R)}(r)=\csc ^2(r)(a^0_4 ((\cos (r)+1) \csc (r))^{-2 \text{j}_L-1}+
b^0_4((\cos (r)+1) \csc (r))^{2 \text{j}_L+1})\nonumber\\
\eea
Again the only non-trivial solution here is $a^0_4=0,b^0_4=0$.\\
For the values of $q_L\ne 0$ in the range $|q_L|<j_L$ the analysis is a bit involved. We start from
\bea
A_3{(j_L,-q_L,-q_R)}(r)&=&\frac{1}{4 q_L}((4 j_L+4 j_L^2-2q_L^2+(-3+2 q_L^2)\cos(2 r))\cot(r)
 A_4{(j_L,-q_L,-q_R)}(r)\nonumber\\
 &-&\cos(r)(5\cos(r)
\partial_rA_4{(j_L,-q_L,-q_R)}(r)
+\sin(r)\partial^2_rA_4{(j_L,-q_L,-q_R)}(r))\nonumber\\
\eea
To get the solution for $A_4{(j_L,-q_L,-q_R)}(r)$ at $r=0$ we plug the following ansatz into the 
kernel equation $\mathcal{E}_3$
\bea
A_4^{(j_L,-q_L,-q_R)}(r)&=r^{\alpha}
\eea
and after trivial rescaling get the indicial equation
\bea
 8(32 j_L^3+16 j_L^4-8j_L(1+\alpha)^2+(1+\alpha)^2(-3+2\alpha+\alpha^2)-8 j_L^2(-1+2\alpha+\alpha^2)))=0
\eea
which is solved to yield
\bea
\alpha=-3-2 j_L,\hspace{.8mm}-1 - 2 j_L,\hspace{.8mm}-1 +2 j_L,\hspace{.8mm}1 + 2 jL
\eea
 and ansatz at $r=\pi-r$
\bea
A_4^{(j_L,-q_L,-q_R)}(r)&=(\pi-r)^{\alpha}
\eea
gives the same indicial equation as above and hence the same solution for $\alpha$. Smooth solutions correspond to
 $\alpha=(2 j_L+1),(2j_L-1)$.\\
 Now in general it is very difficult to solve a fourth order ordinary differential equations. But fortunately in this case it is easy to see that no non-trivial solution exists by using a simple trick:
if the smooth solution at $r=0$ interpolates to smooth solution at $r=\pi$ then we can try to construct out of the
$\mathcal{E}_3$ a positive definite integral which will indicate a contradiction.\\
Let's suppose a general solution exists for $A_4^{(j_L,-q_L,-q_R)}(r)$ and define a function $S(r)$
\bea
A_4^{(j_L,-q_L,-q_R)}(r)&=&F(r)\nonumber\\
S(r) &=& F(r)\bigg( 4 (3+8 j_L+40 j_L^2+64 j_L^3+32 j_L^4-15 q_L^2-32 j_Lq_L^2-32 j_L^2q_L^2\nonumber\\
&+&12 q_L^4+2(-9+14 q_L^2-8 q_L^4+4 j_L(-3+4q_L^2)\nonumber\\
&+&4 j_L^2(-3+4 q_L^2))\cos(2r)+(9-13 q_L^2+4 q_L^4)\cos(4 r))F(r)\nonumber\\
&-&2\sin(r)(4\cos(r)(27+24 j_L+24 j_L^2-20 q_L^2\nonumber\\
&+&10(-3+2 q_L^2)\cos(2 r))\partial_r F(r)+2\sin(r)((-5+16 j_L+16 j_L^2\nonumber\\
&-&8 q_L^2+(-37+8q_L^2)\cos(2 r))\partial^2F(r)-2\sin(r)(10 \cos(r)\partial_r^3F(r)\nonumber\\
&+&\sin(r)\partial^4_rF(r)))))
\bigg).
\eea
We perform integration by parts until the integrand is converted into  a sum of positive terms plus the total derivative terms.
 The total derivative terms becomes the following boundary contributions
\bea
S_{boundary}&=&-2(7+8 j_L+8 j_L^2-4 q_L^2+4(-1+q_L^2)\cos(2 r))F(r)^2\sin(2 r)\nonumber\\
&-&8\sin(r)^3\partial_rF(r)(\cos(r)\partial_rF(r)+\sin(r)\partial^2_rF(r))\nonumber\\
&-&4 F(r)\sin(r)^2((7+16 j_L+16 j_L^2-8 q_L^2+(-13+8q_L^2)\cos(2 r)\partial_rF(r)\nonumber\\
&-&2\sin(r)(6\cos(r)\partial^2_rF(r)+\sin(r)\partial^3_rF(r)).
\eea
We now first check that the boundary term vanishes. By assumption of the smoothness of $F(r)$ it goes at least as
$r^{2j_L-1}$ at $r=0$. This means that $\partial^3_rF(r)F(r)$ and $\partial^2_rF(r)\partial_rF(r)$ goes like $r^{4 j_L-5}$. The coefficient of these terms is $\sin(r)^4$ which goes as $r^4$. Combining this we get $r^{4 j_L-1}$ which vanishes if $j_L\ge \frac{1}{2}$. Similarly all the other terms can be easily seen to vanish at $r=0$ and $r=\pi$.
Next we define another function $S_{bulk}$  related to $S_{boundary}$ as
\bea\label{eq:kernelbulkS1}
S_{bulk}(r)&=&S(r)-\partial_rS_{boundary}\nonumber\\
&=&8\sin(r)^4(\partial^2_rF(r))^2+4(9+16 j_L+16  j_L^2-8 q_L^2\nonumber\\
&+&(-9+8 q_L^2)\cos(2r))\sin(r)^2(\partial_rF(r))^2\nonumber\\
&+&4(3+8 j_L+40 j_L^2+64 j_L^3+32 j_L^4-15 q_L^2-32j_Lq_L^2\nonumber\\
&-&32 j_L^2q_L^2+12 q_L^4+(-11+24q_L^2-16q_L^4+16j_L(-1+2q_L^2)\nonumber\\
&+&16 j_L^2(-1+2 q_L^2))\cos(2 r)+(5-9 q_L^2+4 q_L^4)\cos(4 r)).
\eea
Notice that the coefficient of  $(\partial^2_rF(r))^2$ is positive definite. Similarly in the coefficient of $ (\partial_rF(r))^2$,
$(1-\cos(2 r))$ is non-negative and bounded by $2$ and since $j_L^2>q_L^2$, the coefficient of $ (\partial_rF(r))^2$ is non-negative.\\
Now looking at the coefficient of $F(r)^2$ term , since only even powers of $q_L$ appear we can take $q_L$ to be positive. Since $q_L<j_L$ in units of $1$, we can set $j_L=q_L+n$ for $n=0,1,2,3...$. This coefficient is a function of $r$, so we will find its minimum as a function of $r$ for fixed $q_L$ and $n$. If the value of the coefficient at the minimum is non-negative for the allowed values of $q_L$ and $n$ then this means that $S_{bulk}$ is a sum of nonnegative terms and hence positive definite, which is a contradiction to $\mathcal{E}_3$.\\
\bea
\partial_r\text{Coefficient} (F(r))&=&\sin(2 r)(4(-2(-11+24 q_L^2-16q_L^4+16(1+n+q_L)\nonumber\\
&+&16(1+n+q_L)^2(-1+2 q_L^2))-8(5-9 q_L^2+4 q_L^4)\cos(2r))).\nonumber\\
\eea
For $q_L=1$ it evaluates to $-8(93+80n+16 n^2)$, so one solution is $r=0,\frac{\pi}{2},\pi$ for $\sin(2 r)=0$. \\For 
$q_L\ne 1$ the solution is 
\bea
\cos (2 r)= \frac{8 \left(-2 \text{q}_L^4-52 \text{q}_L^3-169
   \text{q}_L^2+20 \text{q}_L+80\right)+48 \text{q}_L+43}{4 \left(4
   \text{q}_L^4-9 \text{q}_L^2+5\right)}
\eea
It can be easily shown that the absolute value of the right hand side is greater than $1$ and hence the solution does not exist. As an example for $q_L=\frac{1}{2}$
\bea
\cos(2 r)=\frac{8+8 n +2 n^2}{3}>1
\eea
so this stationary point does not exist. \\For genera argument for $q_L>1$ take the numerator and denominator of the general expression for $\cos(2 r)$ separately as
\bea
X_N&=&8 \left(-2 \text{q}_L^4-52 \text{q}_L^3-169
   \text{q}_L^2+20 \text{q}_L+80\right)+48 \text{q}_L+43,\nonumber\\
X_D&=&4 \left(4 \text{q}_L^4-9 \text{q}_L^2+5\right).
\eea
and observe that for $q_L=\frac{3}{2}$, $X_N<0$ and $X_D>0$. Now for all $q_L>1$ i.e. for $q_L=p+1$ for $p$ increasing in increments of $\frac{1}{2}$, let' s perform a Taylor series expansion of $X_N$ around $n=0$
\bea
X_N&=&-16 n^2 \left(2 p^2+4 p+1\right)-16 n \left(4 p^3+18 p^2+22
   p+5\right)\nonumber\\
&-&16 p^4-160 p^3-456 p^2-448 p-93+O(n)^5.
\eea
For $q_L>1$ this is negative. Furthermore the denominator $X_D=4(1-q_L^2)(5-4q_L^2)$ is positive for $q_L>1$ as then 
$q_L\ge \frac{3}{2}$.\\
Next we are going to show that $|X_N|=-X_N>X_D$, or in other words $|X_N|-X_D$ is positive. To show this again perform Taylor series expansion of  $|X_N|-X_D$ around $n=0\quad for \quad q_L=p+1$
\bea
|X_N|-X_D&=&16 n^2 \left(2 p^2+4 p+1\right)+16 n \left(4 p^3+18 p^2+22
   p+5\right)\nonumber\\
&+&96 p^3+396 p^2+456 p+93+O(n)^5.
\eea
It is clear that for $q_L>1$ each of these terms is positive so $\frac{|X_N|}{X_D}>1$ for all $q_L>1$. \\
Therefore $\cos(2 r)=\frac{|X_N|}{X_D}$ has no solution for real $r$. This means that the only stationary point is 
$\sin(2 r)=0$ i.e. for $r=0,\frac{\pi}{2},\pi$. Next taking the second derivative of the coefficient of $F(r)^2$ and evaluating at $r=0,\pi$, for $q_L=\frac{1}{2}$ we get $128 (1 + n) (3 + n)>0$, so stationary point at $r=0,\pi$ are minimum for $q_L=\frac{1}{2}$. To go for $q_L=p+1\ge 1$, now perform Taylor series expansion of the second derivative around $p=0$ we get
\bea
-16 (93 + 16 n (5 + n)) - 128 (55 + 44 n + 8 n^2) p - 
 64 (129 + 8 n (9 + n)) p^2
 - 512 (7 + 2 n) p^3 - 512 p^4+O(p)^5\nonumber\\
\eea
Note that each term is negative, so at $r=0,\pi$ this is maximum for $q_L\ge 1$.\\
Now for $r=\frac{\pi}{2}$ and $q_L=\frac{1}{2}$, evaluating the second derivative we get $-128 (6 + 4 n + n^2)<0$. Next for $q_L=p+1\ge 1$, expand the second derivative in the Taylor series around $r=\frac{\pi}{2},p=0$ to get
\bea
16 (93 + 16 n (5 + n)) + 128 (57 + 44 n + 8 n^2) p + 
 64 (99 + 8 n (9 + n)) p^2 +
 512 (3 + 2 n) p^3+O(p)^5\nonumber\\
\eea
Each term is positive, so stationary point at $r=\frac{\pi}{2}$ is minimum for $q_L\ge 1$. In the final step we evaluate the coefficient of $F(r)^2$ at the minimum and show that it is non-negative in all the cases.\\
First consider $q_L=\frac{1}{2}$ and evaluating the coefficient at the minimum $r=0$
\bea
4 (417 + 928 n + 744 n^2 + 256 n^3 + 32 n^4)>0
\eea
Next consider $q_L\ge 1$, in this  case the minimum is at $r=\frac{\pi}{2}$ and the series expansion of the  coefficient around $n=0$ and $q_L=p+1$ evaluates to
\bea
&12 (305 + 328 p + 88 p^2) + 32 (215 + 206 p + 48 p^2) n + 
 32 (143 + 104 p + 16 p^2) n^2 \nonumber\\
&+ 256 (5 + 2 p) n^3 + 128 n^4+O(n)^7
\eea
showing that each coefficient in this series expansion is positive, so the coefficient of $F(r)^2$ is already positive at the minimum value as a function of $r$.\\
This proves that $S_{bulk}$ is a sum of non-negative terms, therefore each term must vanish, which implies that there is no non-singular solution for $F(r)$. Hence the conclusion is that Kernel of $D_{10}$ is $\bf{empty}$.\\
\end{appendices}
\newpage

\appendix




\bibliographystyle{plain}

\bibliography{bibliography}

\begin{thebibliography}{10}

\bibitem{barut1986theory}
A.~O. Barut and Ryszard Raczka.
\newblock {\em Theory of group representations and applications}.
\newblock World Scientific, Singapore, 1986.

\bibitem{Bullimore:2014nla}
Mathew Bullimore, Martin Fluder, Lotte Hollands, and Paul Richmond.
\newblock {The superconformal index and an elliptic algebra of surface
  defects}.
\newblock {\em JHEP}, 10:62, 2014.

\bibitem{Cabo-Bizet:2016ars}
Alejandro Cabo-Bizet.
\newblock {Factorising the 3D Topologically Twisted Index}.
\newblock {\em JHEP}, 04:115, 2017.

\bibitem{Drukker:2010jp}
Nadav Drukker, Davide Gaiotto, and Jaume Gomis.
\newblock {The Virtue of Defects in 4D Gauge Theories and 2D CFTs}.
\newblock {\em JHEP}, 06:025, 2011.

\bibitem{Gaiotto:2014gha}
Davide Gaiotto.
\newblock {Boundary F-maximization}.
\newblock 2014.

\bibitem{Gaiotto:2008sa}
Davide Gaiotto and Edward Witten.
\newblock {Supersymmetric Boundary Conditions in N=4 Super Yang-Mills Theory}.
\newblock {\em J. Statist. Phys.}, 135:789--855, 2009.

\bibitem{Hama:2012bg}
Naofumi Hama and Kazuo Hosomichi.
\newblock {Seiberg-Witten Theories on Ellipsoids}.
\newblock {\em JHEP}, 1209:033, 2012.

\bibitem{Hama:2011ea}
Naofumi Hama, Kazuo Hosomichi, and Sungjay Lee.
\newblock {SUSY Gauge Theories on Squashed Three-Spheres}.
\newblock {\em JHEP}, 05:014, 2011.

\bibitem{Honda:2013uca}
Daigo Honda and Takuya Okuda.
\newblock {Exact results for boundaries and domain walls in 2d supersymmetric
  theories}.
\newblock {\em JHEP}, 09:140, 2015.

\bibitem{Hori:2013ika}
Kentaro Hori and Mauricio Romo.
\newblock {Exact Results In Two-Dimensional (2,2) Supersymmetric Gauge Theories
  With Boundary}.
\newblock 2013.

\bibitem{Hosomichi:2012ek}
Kazuo Hosomichi, Rak-Kyeong Seong, and Seiji Terashima.
\newblock {Supersymmetric Gauge Theories on the Five-Sphere}.
\newblock {\em Nucl. Phys.}, B865:376--396, 2012.

\bibitem{Floch:2015hwo}
Bruno Le~Floch.
\newblock {S-duality wall of SQCD from Toda braiding}.
\newblock 2015.

\bibitem{Nishioka:2014zpa}
Tatsuma Nishioka and Itamar Yaakov.
\newblock {Generalized indices for $ \mathcal{N} $ = 1 theories in
  four-dimensions}.
\newblock {\em JHEP}, 12:150, 2014.

\bibitem{Pestun:2007rz}
Vasily Pestun.
\newblock {Localization of gauge theory on a four-sphere and supersymmetric
  Wilson loops}.
\newblock {\em Commun. Math. Phys.}, 313:71--129, 2012.

\bibitem{Sugishita:2013jca}
Sotaro Sugishita and Seiji Terashima.
\newblock {Exact Results in Supersymmetric Field Theories on Manifolds with
  Boundaries}.
\newblock {\em JHEP}, 11:021, 2013.

\bibitem{Yoshida:2014ssa}
Yutaka Yoshida and Katsuyuki Sugiyama.
\newblock {Localization of 3d $\mathcal{N}=2$ Supersymmetric Theories on $S^1
  \times D^2$}.
\newblock 2014.

\end{thebibliography}

\end{document}